\documentclass[aps,showpacs,twocolumn]{revtex4-1}
\usepackage{epsfig}
\usepackage{graphicx,color,dcolumn}
\usepackage{epstopdf}
\usepackage{amsmath,amssymb}
\usepackage{multirow}
\usepackage{diagbox}
\usepackage{changes}
\usepackage{booktabs}
\usepackage{threeparttable}
\usepackage{subfigure}
\usepackage{hyperref}
\begin{document}

\title{Exotic $Qq\bar{q}\bar{q}$ states in the chiral quark model}

\author{Yuheng Wu$^1$}\email[E-mail: ]{191002007@njnu.edu.cn}
\author{Ye Yan$^1$}\email[E-mail: ]{221001005@njnu.edu.cn}
\author{Yue Tan$^2$}\email[E-mail: ]{tanyue@ycit.edu.cn (Corresponding author)}
\author{Hongxia Huang$^1$}\email[E-mail: ]{hxhuang@njnu.edu.cn (Corresponding author)}
\author{Jialun Ping$^1$}\email[E-mail: ]{jlping@njnu.edu.cn }
\author{Xinmei Zhu$^3$}\email[E-mail: ]{xmzhu@yzu.edu.cn}
\affiliation{$^1$Department of Physics, Nanjing Normal University, Nanjing, Jiangsu 210097, China}
\affiliation{$^2$Department of Physics, Yancheng Institute of Technology, Yancheng 224000, P. R. China}
\affiliation{$^3$Department of Physics, Yangzhou University, Yangzhou 225009, P. R. China}

\begin{abstract}
In the framework of the chiral quark model, we investigate the $Qq\bar{q}\bar{q}$ ($Q= c, b$ and $q= u, d$) tetraquark system with two structures: $Q\bar{q}$-$q\bar{q}$ and $Qq$-$\bar{q}\bar{q}$. The bound-state calculation shows that for the single channel, there is no evidence for any bound state below the minimum threshold in both $cq\bar{q}\bar{q}$ and $bq\bar{q}\bar{q}$ systems. However, after coupling all channels of two structures, we obtain a bound state below the minimum threshold in the $cq\bar{q}\bar{q}$ system with the energy of $1998$ MeV, and the quantum number is $IJ^{P}=\frac{1}{2}0^{+}$. Meanwhile, in the $bq\bar{q}\bar{q}$ system, two bound states with energies of $5414$ MeV and $5456$ MeV are obtained, and the quantum numbers are $IJ^{P}=\frac{1}{2}0^{+}$ and $IJ^{P}=\frac{1}{2}1^{+}$, respectively.
Besides, we also employe the real-scaling method to search for resonance states in the $cq\bar{q}\bar{q}$ and $bq\bar{q}\bar{q}$ systems. Unfortunately, no genuine resonance states were obtained in both systems. We suggest future experiments to search for these three possible bound states.
\end{abstract}
\maketitle
\section{\label{sec:introduction}Introduction}
It is full of challenges and opportunities to search for exotic states.
So far, many tetraquark and pentaquark states containing heavy quarks have been proposed, such as $T_{cc}$~\cite{LHCb:2021auc}, $X(3872)$~\cite{Belle:2003nnu}, $X(4140)$~\cite{LHCb:2022aki}, $P_{c}(4380)$~\cite{LHCb:2015yax}, and so on. These states allow us to deepen our understanding of nonperturbative quantum chromodynamics(QCD) and hadronic interactions.

In 2016, the D0 Collaboration~\cite{D0:2016mwd} reported a new state $X(5568)$ with a mass and width
\begin{eqnarray}
M_{X(5568)}&=&5568\pm2.9^{+0.9}_{-1.9} ~\mbox{MeV},   \nonumber\\
\Gamma_{X(5568)}&=&21.9\pm6.4^{+5.0}_{-2.5}~ \mbox{MeV}. \nonumber
\end{eqnarray}
It is a good candidate for an exotic tetraquark state because of the four different flavor components (u, d, c and b).
Unfortunately, subsequent experiments from LHCb~\cite{LHCb:2016dxl}, CDF~\cite{CDF:2017dwr}, CMS~\cite{CMS:2017hfy}, and ATLAS~\cite{ATLAS:2018udc} collaborations are not confirmed. On the theoretical side, different approaches have been pursued to investigate this state. In Refs.~\cite{Zanetti:2016wjn,Chen:2016mqt,Wang:2016mee,Agaev:2016lkl}, the authors employed QCD sum rule method to study this state. The $X(5568)$ was explained to be a scalar or axial-vector tetraquark state. In Ref.~\cite{Liu:2016xly}, Liu {\em et al} considered the $X(5568)$ as the near-threshold rescattering effect. In the framework of the effective Hamiltonian approach, Wang and Zhu found a lowest-lying $S$-wave tetraquark state with the flavor $[su][\bar{b}\bar{d}]$, which is about 150 MeV higher than the $X(5568)$~\cite{Wang:2016tsi}. However, some studies have asserted the opposite view. In Refs.~\cite{Huang:2019otd,Guo:2016nhb,Albaladejo:2016eps,Chen:2016npt,Lang:2016jpk,Chen:2016ypj}, the authors thought the reported $X(5568)$ can not be explained as a molecular state or a diqaurk-antidiquark state of $us\bar{d}\bar{b}$.
Besides, the strong decay behavior $X(5568)\rightarrow B_{s}^{0}\pi^{+}$ was also investigated by the QCD sum rule~\cite{Wang:2016wkj,Agaev:2016ijz,Dias:2016dme} and quark model~\cite{Goerke:2016hxf}. The decay width they calculated was in agreement with experimental observations, supporting the existence of $X(5568)$.

Another fully open flavor tetraquark states are $X_{0}(2900)$ and $X_{1}(2900)$ with the quark components $ud\bar{c}\bar{s}$, which were first reported by LHCb Collaboration in the $D^{-}K^{+}$ invariant mass distributions of the decay process $B^{+}\rightarrow D^{+}D^{-}K^{-}$ channel~\cite{LHCb:2020bls,LHCb:2020pxc}. The spin-parity of the two states are $J^{P}=0^{+}$ and $1^{-}$, respectively. Their masses and widths are
\begin{eqnarray}
M_{X_{0}(2900)}&=&2866\pm7 ~\mbox{MeV},   \nonumber\\
\Gamma_{X_{0}(2900)}&=&57\pm13~~\mbox{MeV},   \nonumber\\
M_{X_{1}(2900)}&=&2904\pm5~\mbox{MeV} ,   \nonumber\\
\Gamma_{X_{1}(2900)}&=&110\pm12~ \mbox{MeV}. \nonumber
\end{eqnarray}
These two states have attracted a great deal of interest from theorists, and various interpretations have been proposed. In the framework of quark delocalization color screening model~\cite{Xue:2020vtq}, QCD sum rule~\cite{Agaev:2020nrc}, quasipotential Bethe-Salpeter equation approach~\cite{He:2020btl} and effective Lagrangian approach~\cite{Xiao:2020ltm}, the authors explained $X_{0}(2900)$ as an $S$-wave $\bar{D}^{*}K^{*}$ molecular state. But, in Refs.~\cite{Karliner:2020vsi,Wang:2020xyc}, they interpreted the $X_{0}(2900)$ as an isosinglet compact tetraquark. In Ref.~\cite{Tan:2020cpu}, the authors considered the $X_{0}(2900)$ as a resonance state.
For the $X_{1}(2900)$, in Ref.~\cite{Huang:2020ptc}, Huang {\em et al}  interpret it as a $P$-wave $D^{*}\bar{K}^{*}$ molecule. Refs.~\cite{Chen:2020aos,Zhang:2020oze} argue that $X_{1}(2900)$ is a $P$-wave diquak-antidiquark state. Moreover, the production and decay properties of these states were investigated. In Ref.~\cite{Chen:2020eyu,Burns:2020xne}, they argue that the isospins of $X_{0}(2900)$ and $X_{1}(2900)$ can be investigated by $B\rightarrow DX_{0,1}^{\pm,0}$ decays.

Inspired by the charm (bottom)-strange tetraquark states, as mentioned above, it is natural to investigate the existence of tetraquarks with one heavy and three light quarks.
As is commonly believed, QCD is a fundamental theory of the strong interaction. However, the low energy physics of QCD is much harder to calculate directly from QCD. Various theoretical methods have been proposed to solve this problem, such as Lattice QCD~\cite{Wilson:1974sk}, Quark delocalization color screening model~\cite{Wang:1992wi}, QCD sum rule~\cite{Novikov:1976tn} and so on.

In addition to these methods, the chiral quark model (ChQM) is also a typical approach, which can well describe hadron-hadron interaction~\cite{Valcarce:2005em} and has been successfully employed to explain some tetraquark~\cite{Tan:2022pzi}, pantaquark~\cite{Hu:2021nvs} and hexaquark states~\cite{Pan:2023wrm}.
In this work, we use the chiral quark model to systematically investigate $Qq\bar{q}\bar{q}$ ($Q=c,b$ and $q=u,d$) system with the help of Gaussian expansion method (GEM)~\cite{Hiyama:2003cu}.
GEM is a universal few-body calculation method, which can be used to consider the relative motion between any two quarks, and the Gaussian wave function is used to expand the relative wave function so that the structure of the multiquark system can be obtained. For example, for the tetraquark system, in Ref.~\cite{Tan:2019qwe}, $X(3872)$ was investigated with the help of GEM and the results shown that $X(3872)$ can be described as a mixing state of the dominant charmonium state (70\%) and meson-meson component (30\%); For the pentaquark system, in Ref.~\cite{Hu:2022pae}, $\Omega(2012)$ was suggested to be a $\Xi^{*}\bar{K}$ molecular state with quantum number of $IJ^{P}=0(\frac{3}{2})^{-}$ by the help of the GEM. The calculated distances between quarks confirm the molecular nature of the state; For the dibaryon system, in Ref.~\cite{Pan:2023wrm}, GEM was employed to explore the structure of $d^{*}(2380)$. The radius of $d^{*}(2380)$ was around $0.8$ fm, which indicated that it is a compact hexaquark state.

The structure of this paper is as follows. Section II gives a brief description of the quark model and wave functions. Section III is devoted to the numerical results and discussions. The summary is shown in the last section.

\section{MODEL AND WAVE FUNCTIONS}
\label{wavefunction and chiral quark model}
\subsection{Chiral quark model}
In this paper, the chiral quark
model (ChQM) has been employed to investigate the
$Qq\bar{q}\bar{q}$ ($Q= c, b$ and $q= u, d$) tetraquark system.
The details of the model can be found in Refs.~\cite{10Chen:2018hts,17Yang:2009zzp,18Vijande:2004he}.
Here we just present the Hamiltonian of the model£¬

\begin{eqnarray}
H = \sum_{i=1}^4
\left(m_i+\frac{p_{i}^{2}}{2m_{i}}\right)-T_{CM}+\sum_{j>i=1}^4
\left(V_{ij}^{C}+V_{ij}^{G}+V_{ij}^{\chi}\right).\nonumber\\
\end{eqnarray}
Where $m_{i}$ is the constituent mass of quark(antiquark); $p_{i}$ is the momentum of the quark; $T_{CM}$ is the center-of-mass kinetic energy; $V_{ij}^{C}$ and $V_{ij}^{G}$ are the color confinement and one-gluon-exchange interactions;  $V_{ij}^{\chi}$ ($\chi=\pi,K,\eta,\sigma$) is the Goldstone-boson exchange interaction.

In this work, we focus on the $S$-wave $Qq\bar{q}\bar{q}$ ($Q= c, b$ and $q= u, d$) states, so the tensor force interaction is not included.
For the color confinement, the quadratic form is used here,
\begin{equation}
V_{ij}^{C}= ( -a_{c} r_{ij}^{2}-\Delta) \boldsymbol{\lambda}_i^c \cdot \boldsymbol{\lambda}_j^c .
\end{equation}
Where $a_{c}$ and $\Delta$ are model parameters, and $\boldsymbol{\lambda}_i^c$ is the
color Gell-Mann matrices.

One-gluon exchange potential consists of two parts, coulomb and color-magetism interactions,
\begin{eqnarray}
 V_{ij}^{G}&=& \frac{\alpha_s}{4} \boldsymbol{\lambda}_i^c \cdot \boldsymbol{\lambda}_{j}^c
\left[\frac{1}{r_{ij}}-\frac{2\pi}{3m_im_j}\boldsymbol{\sigma}_i\cdot
\boldsymbol{\sigma}_j
  \delta(\boldsymbol{r}_{ij})\right],   \\
 & &  \delta{(\boldsymbol{r}_{ij})}=\frac{e^{-r_{ij}/r_0(\mu_{ij})}}{4\pi r_{ij}r_0^2(\mu_{ij})},~~ r_{0}(\mu_{ij})=\frac{r_0}{\mu_{ij}}.
  \nonumber
\end{eqnarray}
Where $\mu_{ij}$ is the reduced mass between two quarks; $\boldsymbol{\sigma}$ means
the SU(2) Pauli matrices; $\alpha_{s}$ is an effective scale-dependent running coupling,
\begin{equation}
 \alpha_s(\mu_{ij})=\frac{\alpha_0}{\ln\left[(\mu_{ij}^2+\mu_0^2)/\Lambda_0^2\right]}.
\end{equation}
Goldstone boson exchange comes from the effects of the chiral symmetry spontaneous breaking of QCD in low-energy region.
\begin{eqnarray}
V_{ij}^{\pi}&=& \frac{g_{ch}^2}{4\pi}\frac{m_{\pi}^2}{12m_im_j}
  \frac{\Lambda_{\pi}^2}{\Lambda_{\pi}^2-m_{\pi}^2}m_\pi v_{ij}^{\pi}
  \sum_{a=1}^3 \lambda_i^a \lambda_j^a,  \nonumber \\
V_{ij}^{K}&=& \frac{g_{ch}^2}{4\pi}\frac{m_{K}^2}{12m_im_j}
  \frac{\Lambda_K^2}{\Lambda_K^2-m_{K}^2}m_K v_{ij}^{K}
  \sum_{a=4}^7 \lambda_i^a \lambda_j^a,  \nonumber \\
V_{ij}^{\eta} & = &
\frac{g_{ch}^2}{4\pi}\frac{m_{\eta}^2}{12m_im_j}
\frac{\Lambda_{\eta}^2}{\Lambda_{\eta}^2-m_{\eta}^2}m_{\eta}
v_{ij}^{\eta} \nonumber \\
 && \left[\lambda_i^8 \lambda_j^8 \cos\theta_P
 - \lambda_i^0 \lambda_j^0 \sin \theta_P \right],  \nonumber \\
V_{ij}^{\sigma}&=& -\frac{g_{ch}^2}{4\pi}
\frac{\Lambda_{\sigma}^2}{\Lambda_{\sigma}^2-m_{\sigma}^2}m_\sigma
\left[
 Y(m_\sigma r_{ij})-\frac{\Lambda_{\sigma}}{m_\sigma}Y(\Lambda_{\sigma} r_{ij})\right], \nonumber \\
 v_{ij}^{\chi} & = & \left[ Y(m_\chi r_{ij})-
\frac{\Lambda_{\chi}^3}{m_{\chi}^3}Y(\Lambda_{\chi} r_{ij})
\right]
\boldsymbol{\sigma}_i \cdot\boldsymbol{\sigma}_j, \nonumber \\
& & Y(x)  =   e^{-x}/x .
\end{eqnarray}
Where $Y(x)=e^{-x}/x$ is the standard Yukawa function; $g_{ch}$ is the coupling constant for chiral field, which is determined from the $NN\pi$ coupling constant through
\begin{equation}
\frac{g_{ch}^2}{4\pi}=\left(\frac{3}{5}\right)^2\frac{g_{\pi NN}^2}{4\pi}\frac{m_{u,d}^2}{m_N^2}.
\end{equation}
The other symbols in the above expressions have their usual meanings. All model parameters, which are determined by fitting the meson spectrum, are from the work~\cite{Tan:2020ldi}.

\subsection{The wave-function of $Qq\bar{q}\bar{q}$ system}

\begin{figure}[htp]
  \setlength {\abovecaptionskip} {-0.1cm}
  \centering
  \resizebox{0.50\textwidth}{!}{\includegraphics[width=2.0cm,height=1.0cm]{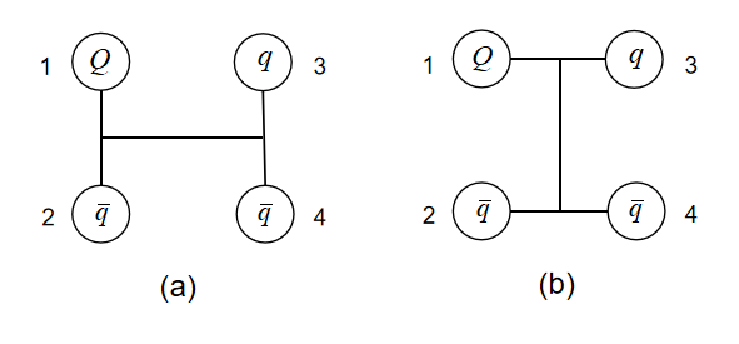}}
  \caption{\label{ab} Configuations of $Qq\bar{q}\bar{q}$ ($Q=c,~b$) tetraquarks.}
\end{figure}

For the $Qq\bar{q}\bar{q}$ system, meson-meson and diquark-antidiquark (replaced by $Q\bar{q}$-$q\bar{q}$ and $Qq$-$\bar{q}\bar{q}$ after) structures are considered, as shown in Fig.~\ref{ab}. Fig.~\ref{ab} (a) represents the $Q\bar{q}$-$q\bar{q}$ structure and (b) means the $Qq$-$\bar{q}\bar{q}$ structure. The wave function of both structures consist of four parts: orbit, spin, flavor and color wave functions. In addition, the wave function of each part is constructed
by coupling two sub-clusters wave functions. Thus, the wave function for each channel will be the tensor product of orbit
($|R_{i}\rangle$), spin ($|S_{j}\rangle$), color ($|C_{k}\rangle$) and flavor ($|F_{l}\rangle$) components,
\begin{equation}\label{bohanshu}
|ijkl\rangle={\cal A} \left[|R_{i}\rangle\otimes|S_{j}\rangle\otimes|C_{k}\rangle\otimes|F_{l}\rangle \right] .
\end{equation}
Where ${\cal A}$ is the antisymmetrization operator. For the $Qq\bar{q}\bar{q}$ system, 
${\cal A}=1-P_{24}$.

For the orbit wave function, the total wave function consists of two sub-clusters orbit wave functions and the relative motion wave function between two sub-clusters.
Since we are interested in $S$-wave system, we put all the orbital angular momenta equal to zero. Here, the magnetic quantum number ($M=0$) is omitted.
\begin{eqnarray}\label{spatialwavefunctions}
|R_{1}\rangle & = & \left[[\Psi_{l_1}({\bf r}_{12})\Psi_{l_2}({\bf
r}_{34})]_{l_{12}}\Psi_{L_r}({\bf r}_{1234}) \right]_{LM_{L}}, \nonumber\\
|R_{2}\rangle&=&\left[[\Psi_{l_1}({\bf r}_{13})\Psi_{l_2}({\bf
r}_{24})]_{l_{12}}\Psi_{L_r}({\bf r}_{1324}) \right]_{LM_{L}}.
\end{eqnarray}
Where the bracket ``[~]" indicates angular momentum coupling, and the ``L" means total orbit angular momentum coupled by $``L_r"$,
relative motion angular momentum, and ``$l_{12}$" coupled by ``$l_1$" and ``$l_2$", sub-cluster angular momenta.
In addition, we use ``$|R_{1}\rangle$" denotes meson-meson structure while ``$|R_{2}\rangle$" denotes diquark-antidiquark structure.
In Gaussian expansion method(GEM), the radial part of spatial wave function is expanded by Gaussians~\cite{Hiyama:2003cu}:
\begin{subequations}
\label{radialpart}
\begin{align}
R(\mathbf{r}) & = \sum_{n=1}^{n_{\rm max}} c_{n}\psi^G_{nlm}(\mathbf{r}),\\
\psi^G_{nlm}(\mathbf{r}) & = N_{nl}r^{l}
e^{-\nu_{n}r^2}Y_{lm}(\hat{\mathbf{r}}),
\end{align}
\end{subequations}
where $N_{nl}$ are normalization constants,
\begin{align}
N_{nl}=\left[\frac{2^{l+2}(2\nu_{n})^{l+\frac{3}{2}}}{\sqrt{\pi}(2l+1)}
\right]^\frac{1}{2}.
\end{align}
$c_n$ are the variational parameters, which are determined dynamically. The Gaussian size parameters are chosen
according to the following geometric progression
\begin{equation}\label{gaussiansize}
\nu_{n}=\frac{1}{r^2_n}, \quad r_n=r_1a^{n-1}, \quad
a=\left(\frac{r_{n_{\rm max}}}{r_1}\right)^{\frac{1}{n_{\rm
max}-1}}.
\end{equation}
This procedure enables optimization of the ranges using just a small number of Gaussians.

For the spin wave functions, there is no difference between quark and antiquark. The meson-meson structure has the same total spin as the diquark-antidiquark structure. The spin wave functions of the cluster are shown below.
\begin{eqnarray}
 & & \chi_{11}^{\sigma}=\alpha\alpha,~~
\chi_{10}^{\sigma}=\frac{1}{\sqrt{2}}(\alpha\beta+\beta\alpha),~~ \nonumber \\
 & & \chi_{1-1}^{\sigma}=\beta\beta,~~\chi_{00}^{\sigma}=\frac{1}{\sqrt{2}}(\alpha\beta-\beta\alpha).
\end{eqnarray}
According to Clebsch-Gordan coefficient table, total spin wave function can be written below.
\begin{eqnarray}
|S_{1}\rangle & = & \chi_{0}^{\sigma1}=\chi_{00}^{\sigma}\chi_{00}^{\sigma}, \nonumber \\
|S_{2}\rangle & = & \chi_{0}^{\sigma2}= \sqrt{\frac{1}{3}}(\chi_{11}^{\sigma}\chi_{1-1}^{\sigma}
  -\chi_{10}^{\sigma}\chi_{10}^{\sigma}+\chi_{1-1}^{\sigma}\chi_{11}^{\sigma}), \nonumber \\
|S_{3}\rangle & = & \chi_{1}^{\sigma1}=\chi_{00}^{\sigma}\chi_{11}^{\sigma}, \\
|S_{4}\rangle & = & \chi_{1}^{\sigma2}=\chi_{11}^{\sigma}\chi_{00}^{\sigma}, \nonumber \\
|S_{5}\rangle & = & \chi_{1}^{\sigma3}=\frac{1}{\sqrt{2}}(\chi_{11}^{\sigma}\chi_{10}^{\sigma}-\chi_{10}^{\sigma}\chi_{11}^{\sigma}),\nonumber \\
|S_{6}\rangle & = & \chi_{2}^{\sigma1}=\chi_{11}^{\sigma}\chi_{11}^{\sigma}. \nonumber
\end{eqnarray}
Where the subscript of ``$\chi_{S}^{\sigma i}$" denotes total spin of the tretraquark, and the superscript is the index of
the spin function with fixed $S$.

The flavor wave functions of the sub-clusters for two structures are shown below,
\begin{eqnarray}
&& \chi_{\frac{1}{2}\frac{1}{2}}^{fm}=Q\bar{d},~~\chi_{\frac{1}{2}-\frac{1}{2}}^{fm}=-Q\bar{u},~~
\chi_{11}^{fm}=u\bar{d},\nonumber\\
&&\chi_{10}^{fm}=\frac{1}{\sqrt{2}}(-u\bar{u}+d\bar{d}),~~
\chi_{00}^{fm}=\frac{1}{\sqrt{2}}(-u\bar{u}-d\bar{d}),~~~~\\
&& \chi_{\frac{1}{2}\frac{1}{2}}^{fd}=Qd,~~\chi_{\frac{1}{2}-\frac{1}{2}}^{fd}=Qu,~~
\chi_{11}^{fd}=\bar{d}\bar{d},\nonumber \\
&&\chi_{00}^{fd}=\frac{1}{\sqrt{2}}\left( \bar{u}\bar{d}-\bar{d}\bar{u} \right),~~
\chi_{10}^{fd}=-\frac{1}{\sqrt{2}}\left( \bar{u}\bar{d}+\bar{d}\bar{u} \right).~~~~
\end{eqnarray}
Where $Q=c,b$; The subscripts of $\chi_{I}^{fm(d)}$ are the isospin and its third component. 
The total flavor wave functions can be written as,
\begin{eqnarray}
|F_{1}\rangle & = & \chi_{\frac{1}{2}}^{fm}=\sqrt{\frac{2}{3}}\chi_{11}^{fm}
 \chi_{\frac{1}{2}-\frac{1}{2}}^{fm}-\sqrt{\frac{1}{3}}\chi_{10}^{fm} \chi_{\frac{1}{2}\frac{1}{2}}^{fm}, \nonumber\\
|F_{2}\rangle & = & \chi_{\frac{1}{2}}^{fm}=\chi_{\frac{1}{2}\frac{1}{2}}^{fm}\chi_{00}^{fm}, \nonumber \\
|F_{3}\rangle & = & \chi_{\frac{3}{2}}^{fm}=\chi_{11}^{fm}  \chi_{\frac{1}{2}\frac{1}{2}}^{fm}, \\
|F_{4}\rangle & = &
\chi_{\frac{1}{2}}^{fd}=\sqrt{\frac{2}{3}}\chi_{11}^{fd}
 \chi_{\frac{1}{2}-\frac{1}{2}}^{fd}-\sqrt{\frac{1}{3}}\chi_{10}^{fd} \chi_{\frac{1}{2}\frac{1}{2}}^{fd}, \nonumber\\
|F_{5}\rangle & = & \chi_{\frac{1}{2}}^{fd}=\chi_{\frac{1}{2}\frac{1}{2}}^{fd}\chi_{00}^{fd}, \nonumber \\
|F_{6}\rangle & = & \chi_{\frac{3}{2}}^{fd}=\chi_{11}^{fd}  \chi_{\frac{1}{2}\frac{1}{2}}^{fd}. \nonumber
\end{eqnarray}
Where the subscript of $\chi_{I}^{fm(d)}$ is total isospin.

The colorless tetraquark system has four color structures, including $1\otimes1$, $8\otimes8$, $\bar{3}\otimes 3$ and $6\otimes \bar{6}$,
\begin{eqnarray}
|C_{1}\rangle & = & \chi_{1\otimes1}^{m1}=\frac{1}{\sqrt{9}}(\bar{r}r\bar{r}r+\bar{r}r\bar{g}g+\bar{r}r\bar{b}b
   +\bar{g}g\bar{r}r+\bar{g}g\bar{g}g \nonumber \\
  & + & \bar{g}g\bar{b}b+\bar{b}b\bar{r}r+\bar{b}b\bar{g}g+\bar{b}b\bar{b}b), \nonumber \\
|C_{2}\rangle & = & \chi_{8\otimes8}^{m2}=\frac{\sqrt{2}}{12}(3\bar{b}r\bar{r}b+3\bar{g}r\bar{r}g+3\bar{b}g\bar{g}b
   +3\bar{g}b\bar{b}g \nonumber \\
 &+ & 3\bar{r}g\bar{g}r+ 3\bar{r}b\bar{b}r+2\bar{r}r\bar{r}r+2\bar{g}g\bar{g}g+2\bar{b}b\bar{b}b-\bar{r}r\bar{g}g \nonumber \\
&-& \bar{g}g\bar{r}r-\bar{b}b\bar{g}g-\bar{b}b\bar{r}r-\bar{g}g\bar{b}b-\bar{r}r\bar{b}b), \\
|C_{3}\rangle & = & \chi^{d1}_{\bar{3}\otimes 3} =\frac{\sqrt{3}}{6}(rg\bar{r}\bar{g}-rg\bar{g}\bar{r}+gr\bar{g}\bar{r}
    -gr\bar{r}\bar{g}+rb\bar{r}\bar{b}, \nonumber \\
&- & rb\bar{b}\bar{r}+br\bar{b}\bar{r}-br\bar{r}\bar{b}+gb\bar{g}\bar{b}-gb\bar{b}\bar{g}+bg\bar{b}\bar{g}-bg\bar{g}\bar{b}), \nonumber \\
|C_{4}\rangle & = & \chi^{d2}_{6\otimes \bar{6}}=\frac{\sqrt{6}}{12}(2rr\bar{r}\bar{r}+2gg\bar{g}\bar{g}+2bb\bar{b}\bar{b}
    +rg\bar{r}\bar{g} \nonumber \\
&+ &rg\bar{g}\bar{r}+gr\bar{g}\bar{r}+gr\bar{r}\bar{g}+rb\bar{r}\bar{b}+rb\bar{b}\bar{r}+br\bar{b}\bar{r} \nonumber \\
&+ &br\bar{r}\bar{b}+gb\bar{g}\bar{b}+gb\bar{b}\bar{g}+bg\bar{b}\bar{g}+bg\bar{g}\bar{b}).\nonumber
\end{eqnarray}
Where $|C_{1}\rangle$, $|C_{2}\rangle$, $|C_{3}\rangle$ and $|C_{4}\rangle$ represent color singlet-singlet ($1\otimes 1$), color octet-octet ($8\otimes8$), color triplet-antitriplet ($\bar{3}\otimes 3$) and color sextet-antisextet ($6\otimes  \bar{6}$) wave functions, respectively. The state with color wave function $|C_{1}\rangle$ is color-singlet channel, and the one with $|C_{2}\rangle$, $|C_{3}\rangle$ or $|C_{4}\rangle$ is hidden-color channel.

Finally, we can acquire the total wave functions by substituting the wave functions of the orbital, the spin, the flavor and the color parts into Eq. (\ref{bohanshu}) according to the definite quantum number of the system.


\section{Result and discussion}
The $S$-wave low-lying states of $Qq\bar{q}\bar{q}$ ($Q= c, b$ and $q= u, d$) tetraquark systems are systematically investigated with both $Q\bar{q}$-$q\bar{q}$ and $Qq$-$\bar{q}\bar{q}$ configurations in the framework of ChQM. The channel coupling of the two configurations are considered. Since we are focus on the $S$-wave states, the orbital angular momentum is set to be zero. The spin quantum number of the $Qq\bar{q}\bar{q}$ system can be $0,~1$ and $2$, so the total angular momentum can be $J=0,~1$ and $2$ for this system. The isospin of the $Q$ ($c,b$) quark is zero. It follows that the isospin of the $Q\bar{q}~(Qq)$ can only be $\frac{1}{2}$, while the isospin of $q\bar{q}~(\bar{q}\bar{q})$ can be $0$ or $1$. In this way, the quantum number of the $Qq\bar{q}\bar{q}$ tetraquark system can be $IJ^{P}=\frac{1}{2}0^{+}, \frac{1}{2}1^{+}, \frac{1}{2}2^{+},\frac{3}{2}0^{+}, \frac{3}{2}1^{+}, \frac{3}{2}2^{+}$.

In the $Qq\bar{q}\bar{q}$ system, we call a color-single state as the scattering state if its energy is above the corresponding threshold. In contrast, a bound state is available if its energy is below the corresponding threshold. For hidden-color channels, which are bound due to their internal color interactions, they can decay to the corresponding color-singlet threshold, possibly forming a color structure resonance state. Besides, if the energy of hidden-color channel is below the minimum color-singlet threshold, it is also a bound state.

\subsection{$cq\bar{q}\bar{q}$ system}
The energies of the $cq\bar{q}\bar{q}$ tetraquark systems for both $c\bar{q}$-$q\bar{q}$ and $cq$-$\bar{q}\bar{q}$ structures, as well as the channel coupling of these two structures are listed in Table~\ref{energy}.

\begin{table*}[htb]
\caption{\label{energy} The energies of the $cq\bar{q}\bar{q}$ system. $F_{i}S_{j}C_{k}$ stands for the index of flavor, spin and color wave functions, respectively. $E_{th}$ means the threshold of corresponding channel, $E_{sc}$ is the energy of every single channel, $E_{cc}$ shows the energy by channel coupling of one certain configuration, and $E_{mix}$ is the lowest energy of the system by coupling all channels of both configurations. (unit: MeV)
}
\begin{tabular}{ccccccc|ccccccccc}\hline\hline
$IJ^{P}$ ~~~&$F_{i}S_{j}C_{k}$ ~~~~&Channel ~~~~&$E_{th}$ ~~~~~&$E_{sc}$ ~~~~&$E_{cc}$ ~~~~&$E_{mix}$~~~~&~~$IJ^{P}$ ~~~&$F_{i}S_{j}C_{k}$ ~~~~&Channel ~~~~&$E_{th}$ ~~~~~&$E_{sc}$ ~~~~&$E_{cc}$ ~~~~&$E_{mix}$~~\\
$\frac{1}{2}0^{+}$ ~~~&111 ~~~~&$\pi D$ ~~~&2002 ~~~~&2004 ~~~&2003 ~~~&1998 ~~~&~~$\frac{1}{2}1^{+}$ ~~~&131 ~~~~&$\pi D^{*}$ ~~~&2119 ~~~~&2121 ~~~&2121 ~~~&2120~ \\
                   &211  ~~~~&$\eta D$   ~~~&2532     ~~~~&2535
                   ~~~&&&&241 ~~~~&$\omega D$  ~~~&2564   ~~~~&2567~~~   \\
                   &221  ~~~~&$\omega D^{*}$   ~~~&2682  ~~~~&2685
                   ~~~&&&&141 ~~~~&$\rho D$   ~~~&2635    ~~~~&2637~~~ \\
                   &121 ~~~~&$\rho D^{*}$  ~~~&2753    ~~~~&2756
                   ~~~&&&&231 ~~~~&$\eta D^{*}$  ~~~&2650 ~~~~&2652~~~ \\
                      &~~             &~~      &~~~ &
                   ~~~&&&&251 ~~~~&$\omega D^{*}$   ~~~&2682  ~~~~&2685~~~   \\
                      &~~             &~~      &~~~ &
                   ~~~&&&&151 ~~~~&$\rho D^{*}$   ~~~&2753   ~~~~&2756~~~   \\

                   &112 ~~~~&$[\pi]_{8}[D]_{8}$ ~~~&     ~~~~&2968 ~~~&
                   ~~~&&&132 ~~~~&$[\omega]_{8} [D]_{8}$    ~~~&  ~~~~&2952 ~~~&~~~ \\
                   &212 ~~~~&$[\eta]_{8}[D]_{8}$    ~~~&  ~~~~&3108  ~~~&
                   ~~~&&&242 ~~~~&$[\rho]_{8} [D]_{8}$   ~~~&  ~~~~&2987 ~~~&~~~\\
                   &222 ~~~~&$[\omega]_{8}[D^{*}]_{8}$ ~~~& ~~~~&2819 ~~~&
                   ~~~&&&142 ~~~~&$[\pi]_{8} [D^{*}]_{8}$    ~~~&   ~~~~&2970 ~~~&~~~ \\
                   &122 ~~~~&$[\rho]_{8} [D^{*}]_{8}$  ~~~&  ~~~~&2812 ~~~&
                   ~~~&&&232 ~~~~&$[\eta]_{8} [D^{*}]_{8}$   ~~~&   ~~~~&3096 ~~~&~~~ \\
                   &~~             &~~      &~~~ &
                   ~~~&&&&252 ~~~~&$[\omega]_{8} [D^{*}]_{8}$   ~~~&  ~~~~&2876 ~~~&~~~ \\
                   &~~             &~~      &~~~ &
                   ~~~&&&&152 ~~~~&   $[\rho]_{8} [D^{*}]_{8}$  ~~~& ~~~~&2908 ~~~&~~~ \\

                   &423 ~~~~&$[cq]_{\bar{3}}[\bar{q}\bar{q}]_{3}$  ~~~&  ~~~~&3037  ~~~&2437 ~
                   ~~~&&&433 ~~~~&$[cq]_{\bar{3}}[\bar{q}\bar{q}]_{3}$  ~~~&  ~~~~&2955  ~~~&2522 ~~~ \\
                   &513 ~~~~&$[cq]_{\bar{3}}[\bar{q}\bar{q}]_{3}$ ~~~&  ~~~&2514
                   ~~~&&&&543 ~~~~&$[cq]_{\bar{3}}[\bar{q}\bar{q}]_{3}$ ~~~& ~~~&2555 ~~~&~~~ \\
                   &414 ~~~~&$[cq]_{6}[\bar{q}\bar{q}]_{\bar{6}}$  ~~~&  ~~~&3054
                   ~~~&&&&453 ~~~~&$[cq]_{\bar{3}}[\bar{q}\bar{q}]_{3}$  ~~~&  ~~~&3017 ~~~&~~~ \\
                   &524 ~~~~&$[cq]_{6}[\bar{q}\bar{q}]_{\bar{6}}$  ~~~&  ~~~&2854
                   ~~~&&&&534 ~~~~&$[cq]_{6}[\bar{q}\bar{q}]_{\bar{6}}$  ~~~& ~~~&3018 ~~~&~~~ \\
                   &~~             &~~      &~~~ &
                   ~~~&&&&554 ~~~~&$[cq]_{6}[\bar{q}\bar{q}]_{\bar{6}}$   ~~~&  ~~~&2935 ~~~&~~~  \\
                   &~~             &~~      &~~~   &
                   ~~~&&&&444 ~~~~&$[cq]_{6}[\bar{q}\bar{q}]_{\bar{6}}$  ~~~&  ~~~&3043 ~~~&~~~  \\ 
                   &~~             &~~      &~~~
                   ~~~&&&&&~~ &~~     &~~~    \\ \hline

$IJ^{P}$ ~~~&$F_{i}S_{j}C_{k}$ ~~~~&Channel ~~~~&$E_{th}$ ~~~~~&$E_{sc}$ ~~~~&$E_{cc}$ ~~~~&$E_{mix}$~~~~&~~$IJ^{P}$ ~~~&$F_{i}S_{j}C_{k}$ ~~~~&Channel ~~~~&$E_{th}$ ~~~~~&$E_{sc}$ ~~~~&$E_{cc}$ ~~~~&$E_{mix}$~~\\
$\frac{1}{2}2^{+}$ ~~~&261 ~~~~&$\omega D^{*}$ ~~~&2682 ~~~~&2685 ~~~&2685 ~~~&2685 ~~~&~~$\frac{3}{2}0^{+}$ ~~~&311 ~~~~&$\pi D$ ~~~&2002 ~~~~&2004 ~~~&2004 ~~~&2004~ \\
                    &161 ~~~~&$\rho D^{*}$  ~~~&2753    ~~~~&2755
                   ~~~&&&&321 ~~~~&$\rho D^{*}$  ~~&2753  ~~~~&2756 ~~~&~~~   \\

                   &262 ~~~~&$[\omega]_{8}[D]_{8}^{*}$   ~~~&  ~~~~&2999   ~~~&
                   ~~~&&&312 ~~~~&$[\pi]_{8} [D]_{8}$   ~~~&  ~~~~&3018 ~~~&~~~ \\
                   &162 ~~~~&$[\rho]_{8} [D^{*}]_{8}$  ~~~&  ~~~~&3059   ~~~&
                   ~~~&&&322 ~~~~&$[\rho]_{8} [D^{*}]_{8}$   ~~~&  ~~~~&2918 ~~~&~~~ \\

                   &463 ~~~~&$[cq]_{\bar{3}}[\bar{q}\bar{q}]_{3}$ ~~~&  ~~~~&2972   ~~~&2972 ~~
                      &&&623 ~~~~&$[cq]_{\bar{3}}[\bar{q}\bar{q}]_{3}$ ~~~&  ~~~~&3054  ~~~&2838 ~~~&~~~ \\
                   &564 ~~~~&$[cq]_{6}[\bar{q}\bar{q}]_{\bar{6}}$  ~~~&  ~~~&3071
                   ~~~&&&&614 ~~~~&$[cq]_{6}[\bar{q}\bar{q}]_{\bar{6}}$  ~~~&   ~~~~&2903 ~~~&~~~\\ 
                      &~~             &~~      &~~~
                      &&&&&~~ &~~     &~~~    \\ \hline

$IJ^{P}$ ~~~&$F_{i}S_{j}C_{k}$ ~~~~&Channel ~~~~&$E_{th}$ ~~~~~&$E_{sc}$ ~~~~&$E_{cc}$ ~~~~&$E_{mix}$~~~~&~~$IJ^{P}$ ~~~&$F_{i}S_{j}C_{k}$ ~~~~&Channel ~~~~&$E_{th}$ ~~~~~&$E_{sc}$ ~~~~&$E_{cc}$ ~~~~&$E_{mix}$~~\\
$\frac{3}{2}1^{+}$ ~~~&331 ~~~~&$\pi D^{*}$ ~~~~&2119 ~~~~&2122 ~~~&2122 ~~~&2122 ~~~&~~$\frac{3}{2}2^{+}$ ~~~&361 ~~~~&$\rho D^{*}$ ~~~&2753 ~~~~&2756   ~~~&2756 ~~~&2756~ \\
                      &341 ~~~~&$\rho D^{~}$  ~~~~&2635  ~~~~&2638
                      ~~~&&&& ~~~&~~     ~        &~~     &~~~    \\
                      &351 ~~~~&$\rho D^{*}$  ~~~~&2753  ~~~~&2756
                      ~~~&&&&&~~             &~~     &~~~ \\

                      &332 ~~~~&$[\pi]_{8}[D^{*}]_{8}$  ~~~&     ~~~~&2998   ~~~&
                      ~~~&&&362 ~~~~&$[\rho]_{8} [D^{*}]_{8}$  ~~~&  ~~~~&3051 ~~~&~~~ \\
                      &342 ~~~~&$[\rho]_{8}[D^{~}]_{8}$  ~~~&  ~~~~&3018  ~~~&
                      ~~~&&&&&~~             &~~     &~~~  \\
                      &352 ~~~~&$[\rho]_{8}[D^{*}]_{8}$  ~~~&   ~~~~&2950 ~~~&
                      ~~~&&&&&~~             &~~     &~~~ \\

                      &633 ~~~~&$[cq]_{\bar{3}}[\bar{q}\bar{q}]_{3}$  ~~~&  ~~~~&2952  ~~~&2875
                      ~~~&&&663 ~~~~&$[cq]_{\bar{3}}[\bar{q}\bar{q}]_{3}$  ~~~&  ~~~~&3036  ~~~&3036 ~~~&~~~ \\
                      &653 ~~~~&$[cq]_{\bar{3}}[\bar{q}\bar{q}]_{3}$  ~~~&  ~~~~&2955
                      ~~~&&&~~ &~~    &~~~ \\

                      &644 ~~~~&$[cq]_{6}[\bar{q}\bar{q}]_{\bar{6}}$  ~~~&  ~~~~&3043
                      ~~~&&&~~ &~~    &~~~ \\

\hline\hline
\end{tabular}
\end{table*}

For the $IJ^{P}=\frac{1}{2}0^{+}$ system, there are eight channels (four color singlet-singlet channels and four color octet-octet channels) of $c\bar{q}$-$q\bar{q}$ structure and four channels (two color triplet-antitriplet channels and two sextet-antisextet channels) of $cq$-$\bar{q}\bar{q}$ structure. From Talbe~\ref{energy} we can see that the energy of each single-channel is above the corresponding theoretical threshold. When we couple eight channels of $c\bar{q}$-$q\bar{q}$ structure, the lowest energy of the system is still above the lowest threshold of the channel $\pi D$.

For the $cq$-$\bar{q}\bar{q}$ structure, the energy of each channel is several hundred MeVs higher than the minimum color single channel ($\pi D$). By coupling with these four channels, the energy of $2437$ MeV is obtained, which is still much higher than the minimum threshold ($\pi D$). These results are qualitatively consistent with those in Ref.~\cite{Lu:2020qmp}, the work of which study the $nn\bar{n}\bar{c}$ (n=u, d) system with the $nn$-$\bar{n}\bar{c}$ structure in the framework of an extended relativized quark model. However, quantitatively, the energy obtained in this work is lower than that in Ref.~\cite{Lu:2020qmp}.
On the one hand, we add the Goldstone boson exchanges interaction terms, which provide attractive interactions in the $cq$-$\bar{q}\bar{q}$ system. On the other hands, the lowest energy of this system is obtained by coupling four channels, as listed in Table~\ref{energy}, while in Ref.~\cite{Lu:2020qmp}, the minimum energy is obtained by coupling two channels, which are ($\mid[nn]_{0}^{\bar{3}}(\bar{n}\bar{c}_{0}^{3})\rangle_{0}$ and $\mid[nn]_{1}^{\bar{6}}(\bar{n}\bar{c}_{1}^{6})\rangle_{0}$).

Then, the channel-coupling is calculated for all the twelve channels and the energy of $1988$ MeV is obtained, which is $4$ MeV lower than the minimum threshold (2002 MeV). This means channel-coupling effects are important for the formation of bound states.
Besides, we also calculate the contributions of each terms in the Hamilton and the root-mean-square (rms) distances in the $IJ^{P}=\frac{1}{2}0^{+}$ system, which are listed in Table~\ref{ECrms1}. The energy contribution values in the table come from the difference between the contribution of each term in the tetraquark system and the sum of its contributions in the two individual mesons. From table~\ref{ECrms1}, we can see that the kinetic and $\eta$-meson exchange terms provide repulsive interactions while the other terms provide attractive interactions. However, the attractive contribution is larger than the repulsive one, which provides the conditions for the $IJ^{P}=\frac{1}{2}0^{+}$ system to form a bound state. Moreover, for the bound state at the energy $1998$ MeV, the rms distances among the quarks are $1.8$-$2.5$ fm and it is composed of $98\%$ $\pi D$, which indicates that this state should be a molecular structure.

\begin{table*}[htb]
\caption{\label{ECrms1} Contributions of all potentials to the binding energy (unit: MeV) and root-mean-square distances (unit: fm) in $cq\bar{q}\bar{q}$ system.}
\begin{tabular}{ccccccccccccccccccccc}\hline\hline
 $IJ^{P}$&~~~energy&~~~&Kinetic ~~~&Conf ~~~~&OGE ~~~~&$\pi$ ~~~~~&$\eta$~~~~~&$\sigma$ ~~~~~&B.E ~~~~~&$r_{12}$ ~~~~~&$r_{13}$ ~~~~~~&$r_{14}$ ~~~~~&$r_{23}$ ~~~~~&$r_{24}$ ~~~~~&$r_{34}$~ \\ \hline
$\frac{1}{2}0^{+}$&~~1998&~~~&50.8 ~~~&-1.9 ~~~~&-13.6 ~~~~&-23.4 ~~~~&1.0 ~~~~&-16.9 ~~~~&-4 ~~~~&1.83 ~~~~&2.52 ~~~~~&1.83 ~~~~~&1.85 ~~~~~&2.56 ~~~~~&1.85 ~\\
\hline\hline
\end{tabular}
\end{table*}

For the $IJ^{P}=\frac{1}{2}1^{+}$ system, 
the energy of each single channel is higher than the corresponding threshold. $c\bar{q}$-$q\bar{q}$ structure and $cq$-$\bar{q}\bar{q}$ structure are coupled separately, and the energies of both structures are $2121$ and $2522$ MeV, which are above the minimum threshold (2119 MeV). Then, channel coupling of two structures has been performed and the energy $E_{mix}=2120$ MeV is obtained, which is still higher than the threshold. Therefore, no bound state below the minimum threshold is found for the $IJ^{P}=01^{+}$ system.

For the $IJ^{P}=\frac{1}{2}2^{+}$ and $IJ^{P}=\frac{3}{2}0^{+}$ systems, both of them have four $c\bar{q}$-$q\bar{q}$ channels and two $cq$-$\bar{q}\bar{q}$ channels. Neither the single channel nor the channel-coupling energies are below the corresponding threshold. Therefore, for both systems, no bound states below the minimum threshold exist.

For the $IJ^{P}=\frac{3}{2}1^{+}$ system, there are nine channels, of which three are color-singlet, three are color-octet, and the remaining three are diquark-antidiquark channels. The single-channel and the channel-coupling calculations tell us that no bound states fall below the minimum threshold.

For the $IJ^{P}=\frac{3}{2}2^{+}$ system, the energy of each single channel is above the threshold of the $\rho D^{*}$. The channel coupling can not help too much. So there is no bound state below the threshold (2753 MeV) for this system, either.

According to the above discussion, there is only one bound state with the binding energy $-4$ MeV for the $IJ^{P}=\frac{1}{2}0^{+}$ in the $cq\bar{q}\bar{q}$ system.
However, it is possible for the hidden-color channels to be resonance states, because the colorful subclusters cannot fall apart directly due to the color confinement. To check the possibility, we carry out

a stabilization method, also named as a real-scaling method, which has proven to be a valuable tool for estimating the metastable energies of electron-atom, electron-molecule, and atom-diatom complexes~\cite{JSimons}.

\begin{figure}[htp]
  \setlength {\abovecaptionskip} {-0.1cm}
  \centering
  \resizebox{0.50\textwidth}{!}{\includegraphics[width=2.0cm,height=1.5cm]{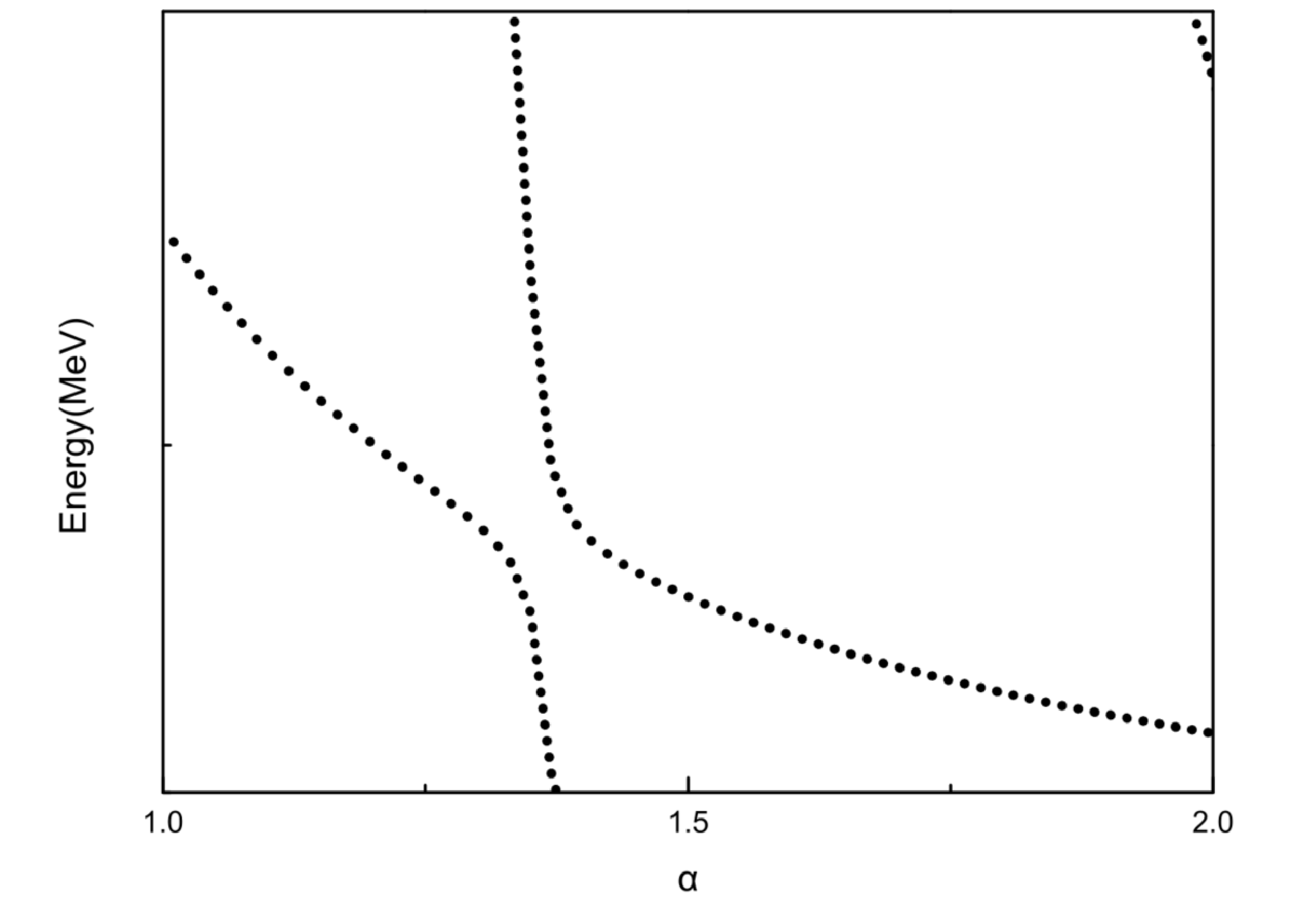}}
  \caption{\label{rsm} The shape of the resonance in the real-scaling method. Taken from Ref.~\cite{JSimons}.}
\end{figure}

In this approach, a factor $\alpha$ is used to scale the finite volume. As $\alpha$ increasing, the false resonances will decay into the corresponding threshold channels, while the genuine resonances repeatedly appear as avoid-crossing structure (as shown in Fig.~\ref{rsm}). This method has been successfully applied to the tetraquark system~\cite{Tan:2022pzi}, petaquark system~\cite{Hiyama:2018ukv}, and so on. It is important to note that for a genuine resonance state, its avoid-crossing structure is formed by a resonance line and a scattering line. However, if there are a large number of coupled channels, the avoid-crossing structure is also formed due to the different rates of the scattering channel descending to the threshold line. We can further estimate whether it is a genuine resonance state by calculating the rms radius. It is important to note that for scattering states it is not square integrable in infinite space, but our calculations are performed in finite space, so a rms distance can be obtained. However, the rms distance of the scattering state will change with increasing space, while the rms distance of the resonance state will remain constant. By calculating the rms distance and the composition, we can estimate whether the avoid-crossing structure is a genuine resonance state.

In this work, the value of $\alpha$ ranges from 1 to 3 to see if there is any resonance state. The results of the $cq\bar{q}\bar{q}$ tetraquark systems with $IJ^{P}=\frac{1}{2}0^{+}, \frac{1}{2}1^{+}, \frac{1}{2}2^{+},\frac{3}{2}0^{+}, \frac{3}{2}1^{+}, \frac{3}{2}2^{+}$ are shown in Figs.~3-8, respectively. We mark the threshold in the red horizontal line and the genuine resonance state or a bound state in the blue horizontal line.

\begin{figure}[htp]
  \setlength {\abovecaptionskip} {-0.1cm}
  \centering
  \resizebox{0.50\textwidth}{!}{\includegraphics[width=2.0cm,height=1.5cm]{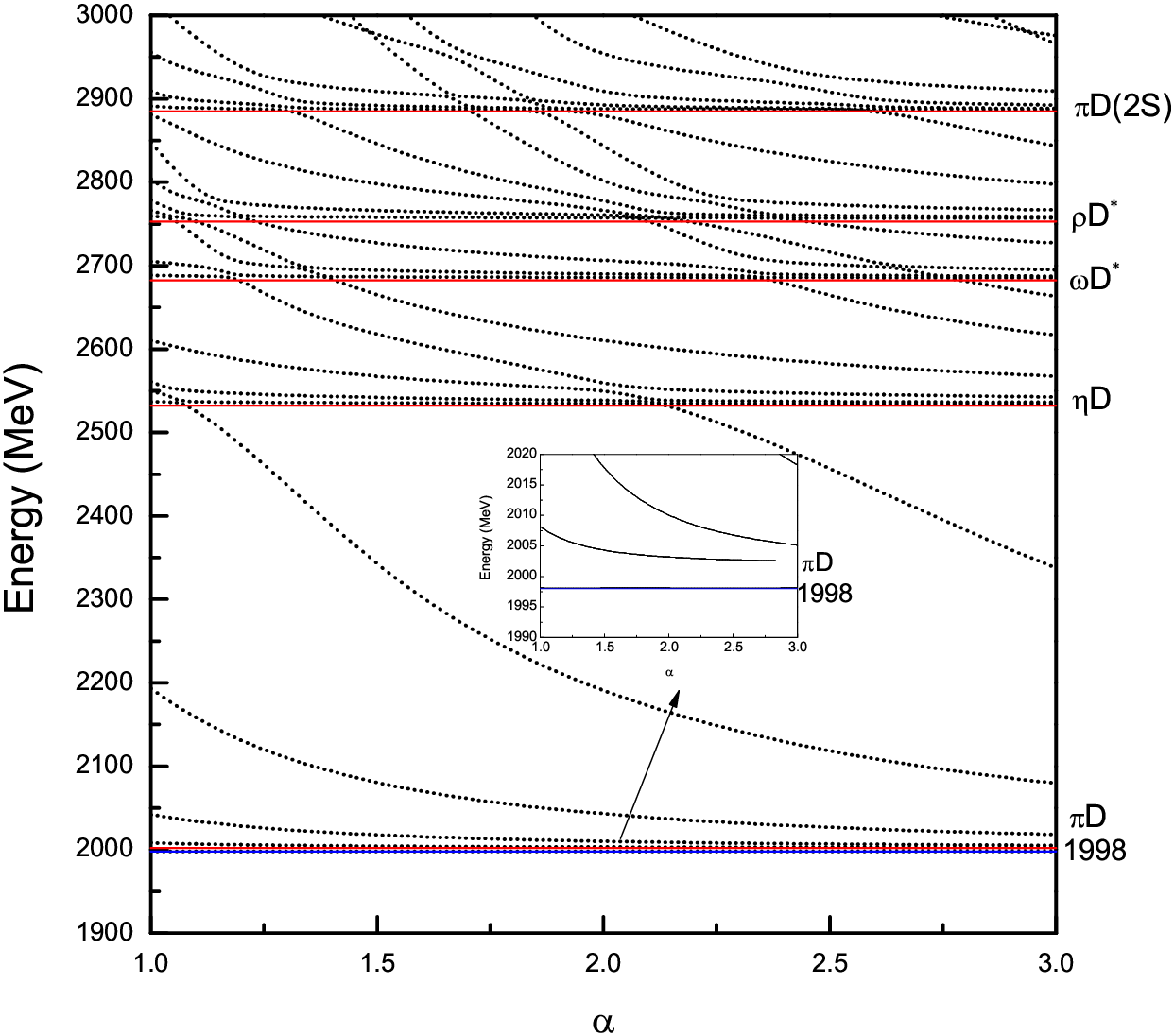}}
  \caption{\label{c1} The stabilization plots of the energies of the $cq\bar{q}\bar{q}$ system with $IJ^{P}=\frac{1}{2}0^{+}$.}
\end{figure}

For the $cq\bar{q}\bar{q}$ system with $IJ^{P}=\frac{1}{2}0^{+}$ in Fig.~\ref{c1}, it is clear that the red horizontal lines locate at the corresponding physical threshold of five channels $\pi D$, $\eta D$, $\omega D^{*}$, $\rho D^{*}$ and $\pi D(2S)$. The blue horizontal line is the bound state at the energy $1998$ MeV. Near energies $2780$ MeV and $2926$ MeV, the avoid-crossing structure is repeated. However, their main components are scattering states (more than 80\%) and the root mean square (rms) distance will larger than $6$ fm with the expansion of space. So, we conclude that both of them are flase resonance states. In this way, there is no resonance state for this system.

\begin{figure}[htp]
  \setlength {\abovecaptionskip} {-0.1cm}
  \centering
  \resizebox{0.50\textwidth}{!}{\includegraphics[width=2.0cm,height=1.5cm]{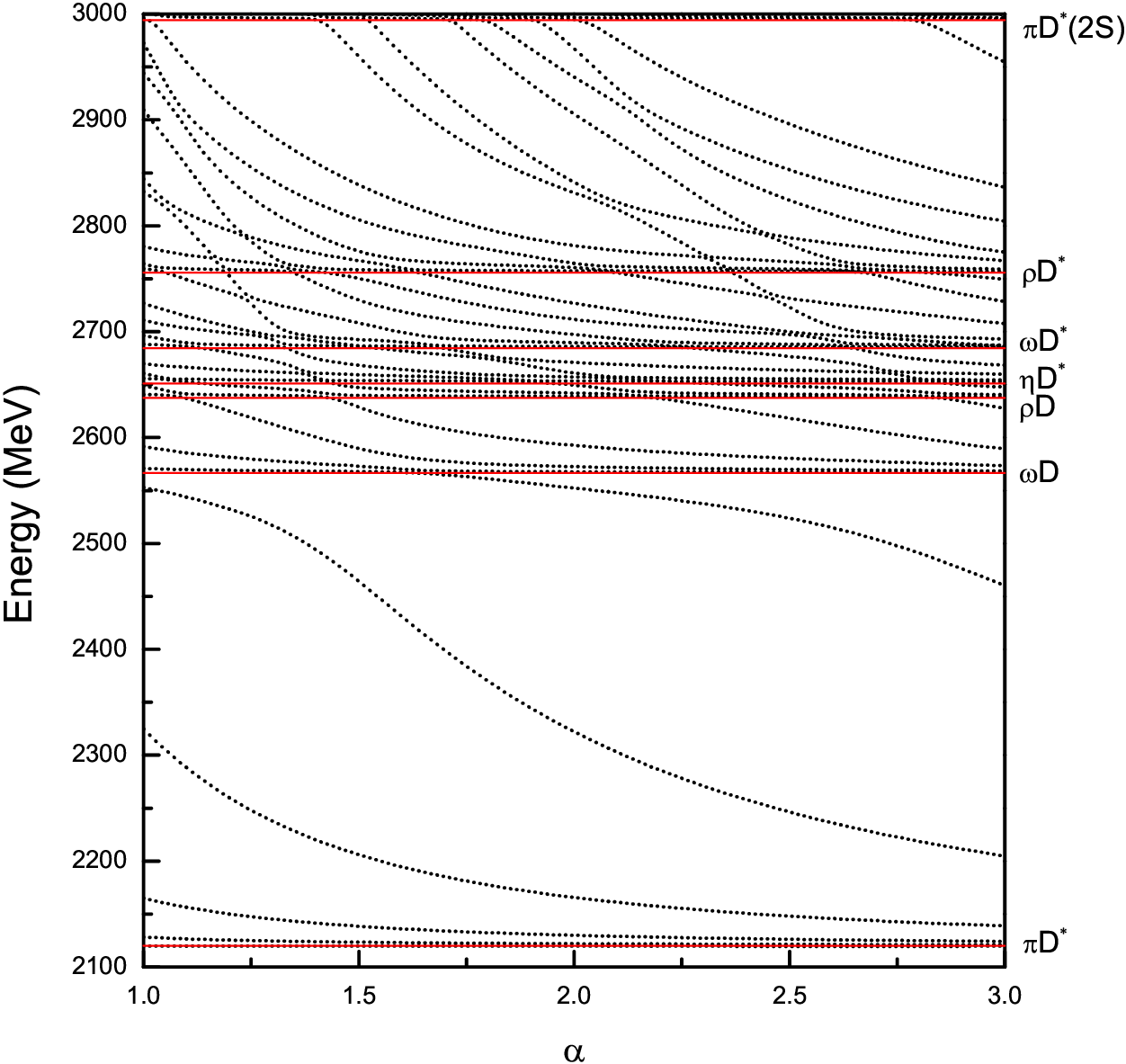}}
  \caption{\label{c2} The stabilization plots of the energies of the $cq\bar{q}\bar{q}$ system with $IJ^{P}=\frac{1}{2}1^{+}$.}
\end{figure}

For the $IJ^{P}=\frac{1}{2}1^{+}$ system, in Fig.~\ref{c2},
seven red horizontal lines from bottom to top represent the thresholds of channels $\pi D^{*}$, $\omega D$, $\rho D$, $\eta D^{*}$, $\omega D^{*}$, $\rho D^{*}$ and $\pi D^{*}(2S)$, respectively. The situation is similar to $IJ^{P}=\frac{1}{2}0^{+}$ system, so there is no genuine resonance state in this system.

\begin{figure}[htp]
  \setlength {\abovecaptionskip} {-0.1cm}
  \centering
  \resizebox{0.50\textwidth}{!}{\includegraphics[width=2.0cm,height=1.5cm]{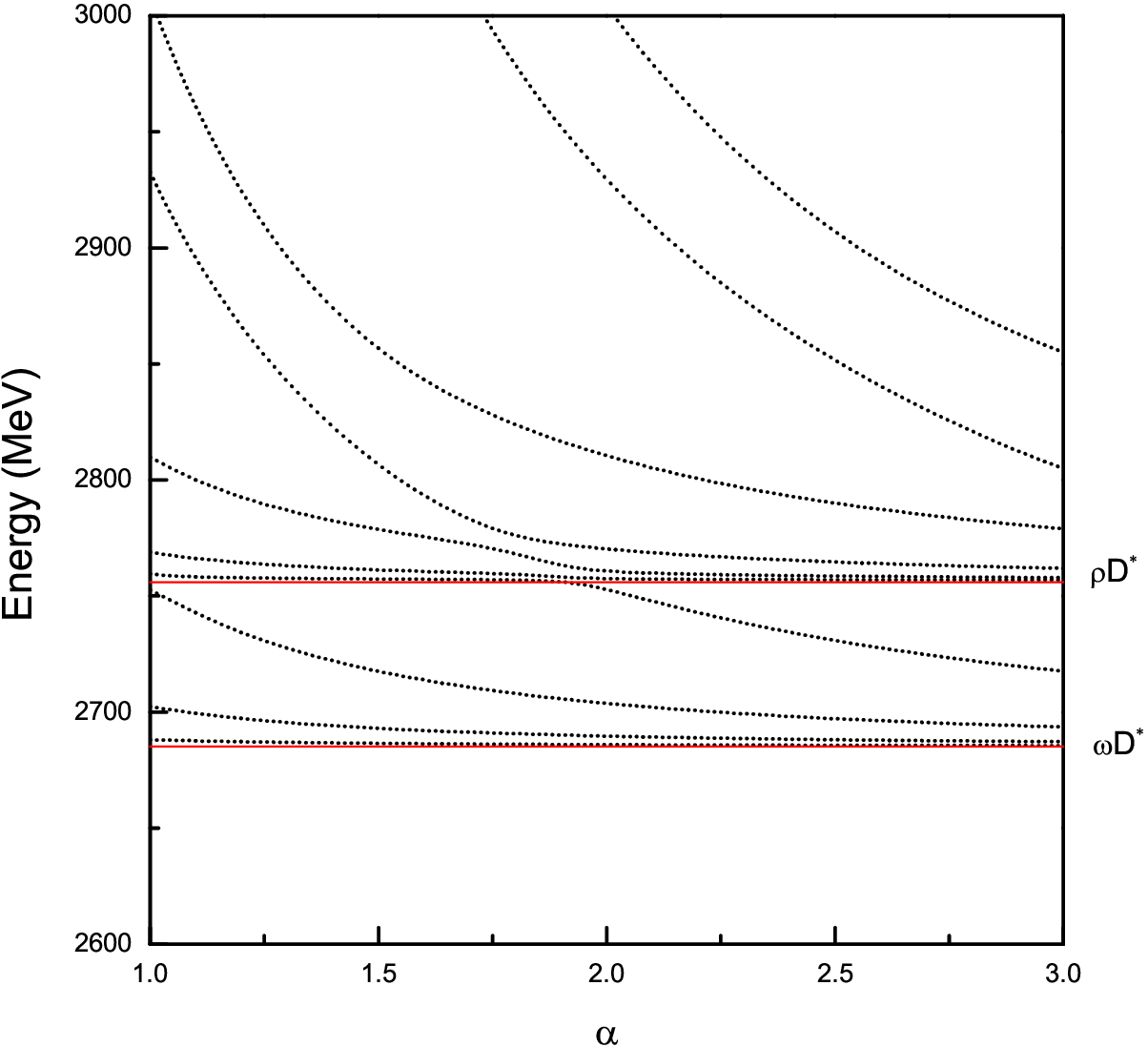}}
  \caption{\label{c3} The stabilization plots of the energies of the $cq\bar{q}\bar{q}$ system with $IJ^{P}=\frac{1}{2}2^{+}$.}
\end{figure}

For the $IJ^{P}=\frac{1}{2}2^{+}$ system, in Fig.~\ref{c3}, two red horizontal lines represent the thresholds of channels $\omega D^{*}$ and $\rho D$, respectively. It is clear that as the $\alpha$ increases, the energy of the continuum state falls towards its threshold. So, there is no resonance state for this system.

\begin{figure}[htp]
  \setlength {\abovecaptionskip} {-0.1cm}
  \centering
  \resizebox{0.50\textwidth}{!}{\includegraphics[width=2.0cm,height=1.5cm]{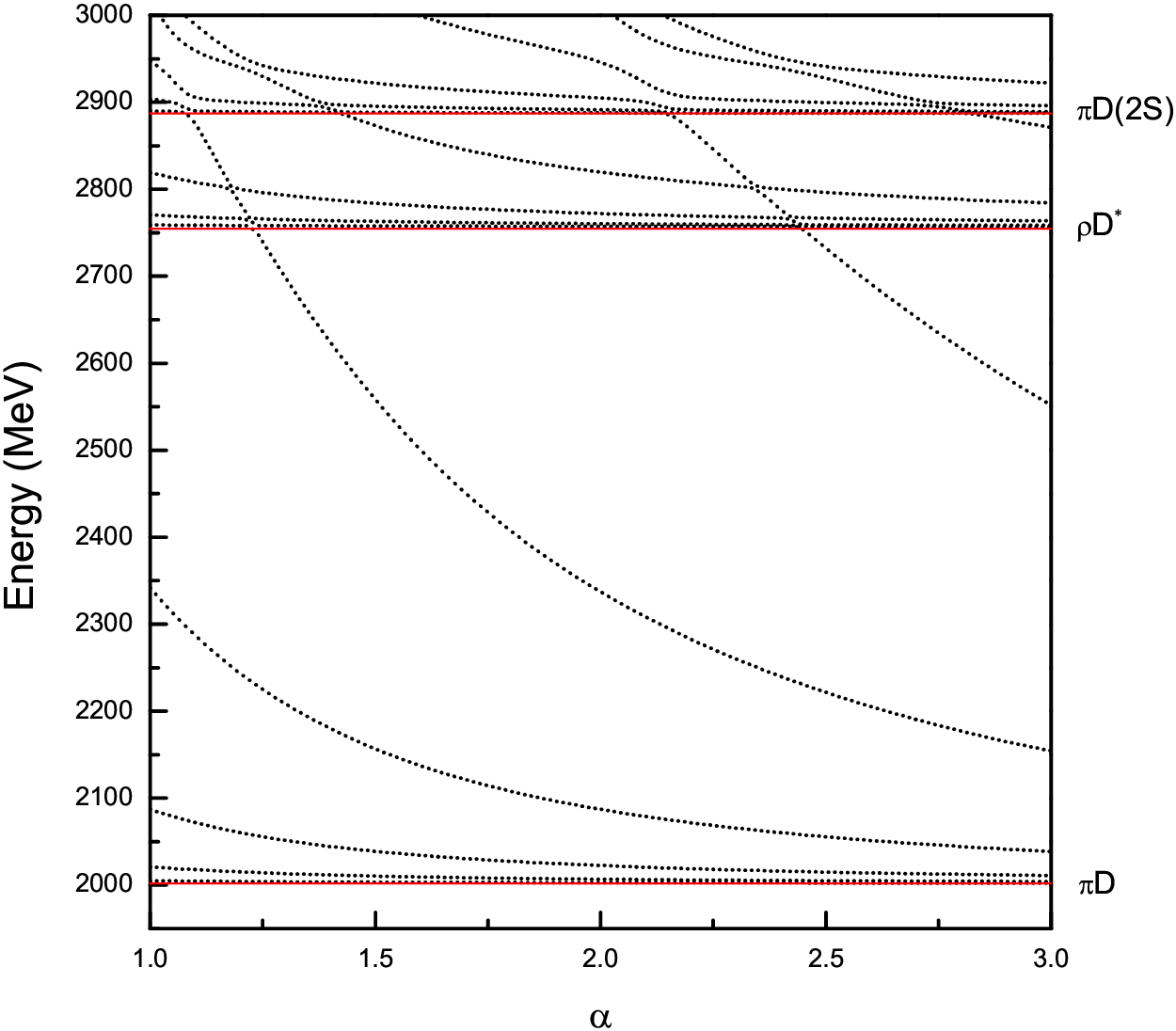}}
  \caption{\label{c4} The stabilization plots of the energies of the $cq\bar{q}\bar{q}$ system with $IJ^{P}=\frac{3}{2}0^{+}$.}
\end{figure}

For the $IJ^{P}=\frac{3}{2}0^{+}$ system, in Fig.~\ref{c4}, the thresholds of channels $\pi D$, $\rho D^{*}$ and $\pi D(2S)$, are marked with red horizontal lines.
In the vicinity of energies $2800$ MeV and $2950$ MeV, the scattering state composition exceeds $95\%$ and $70\%$, respectively. Moreover, their rms distances are larger than $6$ fm with the expansion of space, so that both of them are false resonance states. Therefore, there is no resonance state in this system.

\begin{figure}[htp]
  \setlength {\abovecaptionskip} {-0.1cm}
  \centering
  \resizebox{0.50\textwidth}{!}{\includegraphics[width=2.0cm,height=1.5cm]{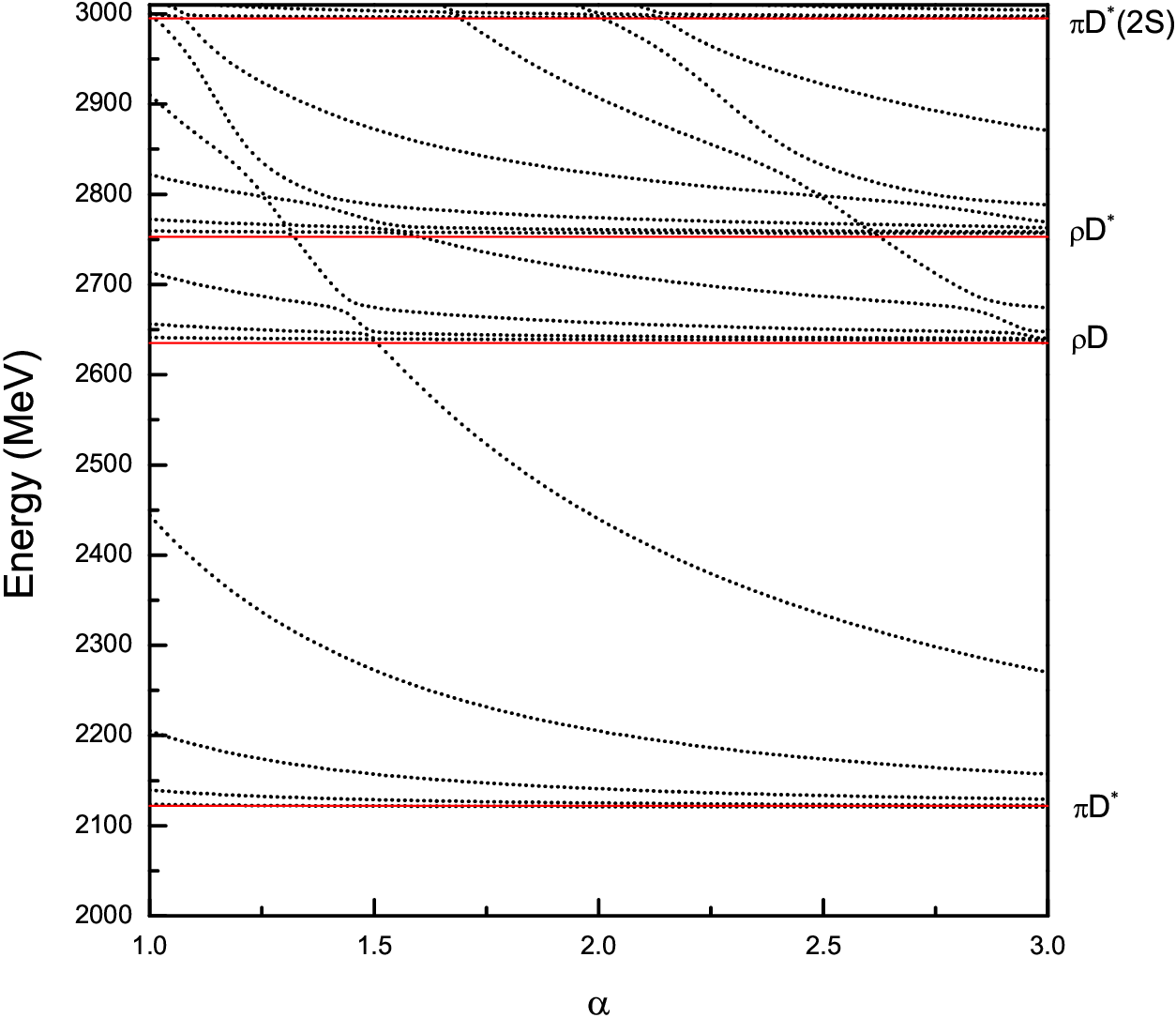}}
  \caption{\label{c5} The stabilization plots of the energies of the $cq\bar{q}\bar{q}$ system with $IJ^{P}=\frac{3}{2}1^{+}$.}
\end{figure}

For the $IJ^{P}=\frac{3}{2}1^{+}$ system, in Fig.~\ref{c5}, the thresholds of channels $\pi D^{*}$, $\rho D$, $\rho D^{*}$ and $\pi D^{*}(2S)$ are marked with red horizontal lines.
Around the energy $2673$ MeV and $2800$ MeV, there are avoid-crossing structures. However, their rms distances are also unstable with increasing space, and their main components are also scattering states (around $93\%$ and $86\%$, respectively),
thus both of them are false resonance states. Therefore, there are also no resonance states in this system.

\begin{figure}[htp]
  \setlength {\abovecaptionskip} {-0.1cm}
  \centering
  \resizebox{0.50\textwidth}{!}{\includegraphics[width=2.0cm,height=1.5cm]{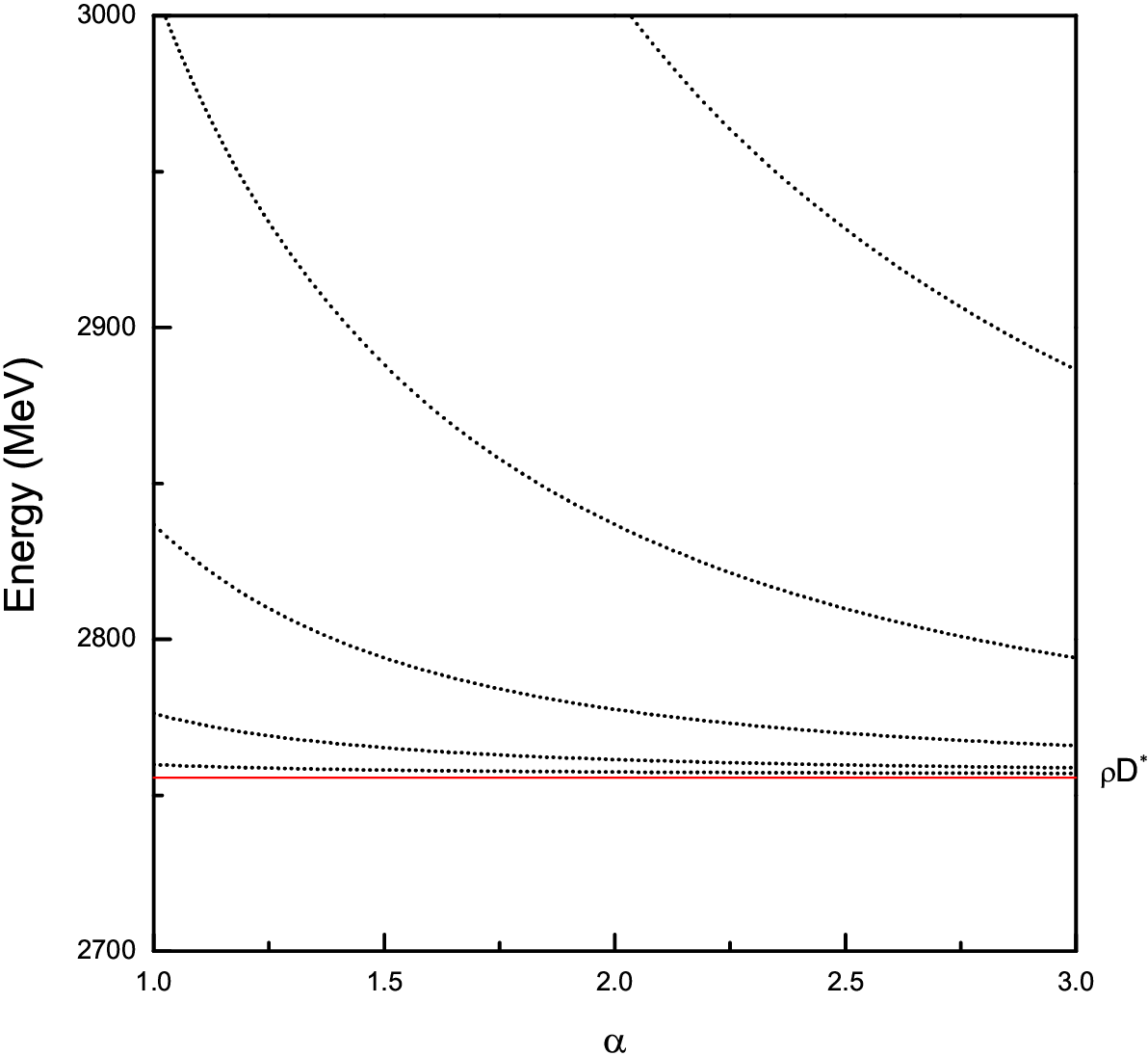}}
  \caption{\label{c6} The stabilization plots of the energies of the $cq\bar{q}\bar{q}$ system with $IJ^{P}=\frac{3}{2}2^{+}$.}
\end{figure}

For the $IJ^{P}=\frac{3}{2}2^{+}$ system, in Fig.~\ref{c6}, the red horizontal line represents the threshold of the channel $\rho D^{*}$. The case is similar to the $IJ^{P}=\frac{1}{2}2^{+}$ system, so there is no resonance state in the $IJ^{P}=\frac{3}{2}1^{+}$ system.

\subsection{$bq\bar{q}\bar{q}$ system}

The energies of the $bq\bar{q}\bar{q}$ tetraquark system are listed in Table~\ref{energy2}. 
Meson-meson structure, diquark-antidiquark structure and channel-coupling of two configurations are considered. Here, we also focus on the $S$- wave state. So, the possible quantum numbers are the same as the $cq\bar{q}\bar{q}$ system.
Since the specific analysis is similar to that of $cq\bar{q}\bar{q}$ system, to save space, we only give a brief description of the results for $bq\bar{q}\bar{q}$ system.
\begin{table*}[htb]
\caption{\label{energy2} The energies of the $bq\bar{q}\bar{q}$ system. $F_{i}S_{j}C_{k}$ stands for the index of flavor, spin and color wave functions, respectively. $E_{th}$ means the threshold of corresponding channel, $E_{sc}$ is the energy of every single channel, $E_{cc}$ shows the energy by channel coupling of one certain configuration, and $E_{mix}$ is the lowest energy of the system by coupling all channels of both configurations. (unit: MeV)}
\begin{tabular}{ccccccc|ccccccccc}\hline\hline
$IJ^{P}$ ~~~&$F_{i}S_{j}C_{k}$ ~~~~&Channel ~~~~&$E_{th}$ ~~~~~&$E_{sc}$ ~~~~&$E_{cc}$ ~~~~&$E_{mix}$~~~~&~~$IJ^{P}$ ~~~&$F_{i}S_{j}C_{k}$ ~~~~&Channel ~~~~&$E_{th}$ ~~~~~&$E_{sc}$ ~~~~&$E_{cc}$ ~~~~&$E_{mix}$~~\\
$\frac{1}{2}0^{+}$ ~~~&111 ~~~~&$\pi B$ ~~~&5419 ~~~~&5421 ~~~&5420 ~~~&5414 ~~~&~~$\frac{1}{2}1^{+}$ ~~~&131 ~~~~&$\pi B^{*}$ ~~~&5458 ~~~~&5460 ~~~&5459 ~~~&5456~ \\
                   &211  ~~~~&$\eta B$   ~~~&5950     ~~~~&5952
                   ~~~&&&&241 ~~~~&$\omega B$  ~~~&5982   ~~~~&5985~~~   \\
                   &221  ~~~~&$\omega B^{*}$   ~~~&6021  ~~~~&6023
                   ~~~&&&&141 ~~~~&$\rho B$   ~~~&6053    ~~~~&6055~~~ \\
                   &121 ~~~~&$\rho B^{*}$  ~~~&6092    ~~~~&6094
                   ~~~&&&&231 ~~~~&$\eta B^{*}$  ~~~&5989 ~~~~&5991~~~ \\
                      &~~             &~~      &~~~ &
                   ~~~&&&&251 ~~~~&$\omega B^{*}$   ~~~&6021  ~~~~&6023~~~   \\
                      &~~             &~~      &~~~ &
                   ~~~&&&&151 ~~~~&$\rho B^{*}$   ~~~&6092   ~~~~&6094~~~   \\

                   &112 ~~~~&$[\pi]_{8}[B]_{8}$ ~~~&     ~~~~&6327 ~~~&
                   ~~~&&&132 ~~~~&$[\omega]_{8} [B]_{8}$    ~~~&  ~~~~&6310 ~~~&~~~ \\
                   &212 ~~~~&$[\eta]_{8}[B]_{8}$    ~~~&  ~~~~&6465  ~~~&
                   ~~~&&&242 ~~~~&$[\rho]_{8} [B]_{8}$   ~~~&  ~~~~&6342 ~~~&~~~\\
                   &222 ~~~~&$[\omega]_{8}[B^{*}]_{8}$ ~~~& ~~~~&6207 ~~~&
                   ~~~&&&142 ~~~~&$[\pi]_{8} [B^{*}]_{8}$    ~~~&   ~~~~&6328 ~~~&~~~ \\
                   &122 ~~~~&$[\rho]_{8} [B^{*}]_{8}$  ~~~&  ~~~~&6213 ~~~&
                   ~~~&&&232 ~~~~&$[\eta]_{8} [B^{*}]_{8}$   ~~~&   ~~~~&6461 ~~~&~~~ \\
                   &~~             &~~      &~~~ &
                   ~~~&&&&252 ~~~~&$[\omega]_{8} [B^{*}]_{8}$   ~~~&  ~~~~&6258 ~~~&~~~ \\
                   &~~             &~~      &~~~ &
                   ~~~&&&&152 ~~~~&   $[\rho]_{8} [B^{*}]_{8}$  ~~~& ~~~~&6279 ~~~&~~~ \\

                   &423 ~~~~&$[bq]_{\bar{3}}[\bar{q}\bar{q}]_{3}$  ~~~&  ~~~~&6383  ~~~&5828 ~
                   ~~~&&&433 ~~~~&$[bq]_{\bar{3}}[\bar{q}\bar{q}]_{3}$  ~~~&  ~~~~&6318  ~~~&5858 ~~~ \\
                   &513 ~~~~&$[bq]_{\bar{3}}[\bar{q}\bar{q}]_{3}$ ~~~&  ~~~&5877
                   ~~~&&&&543 ~~~~&$[bq]_{\bar{3}}[\bar{q}\bar{q}]_{3}$ ~~~& ~~~&5892 ~~~&~~~ \\
                   &414 ~~~~&$[bq]_{6}[\bar{q}\bar{q}]_{\bar{6}}$  ~~~&  ~~~&6415
                   ~~~&&&&453 ~~~~&$[bq]_{\bar{3}}[\bar{q}\bar{q}]_{3}$  ~~~&  ~~~&6359 ~~~&~~~ \\
                   &524 ~~~~&$[bq]_{6}[\bar{q}\bar{q}]_{\bar{6}}$  ~~~&  ~~~&6256
                   ~~~&&&&534 ~~~~&$[bq]_{6}[\bar{q}\bar{q}]_{\bar{6}}$  ~~~& ~~~&6378 ~~~&~~~ \\
                   &~~             &~~      &~~~ &
                   ~~~&&&&554 ~~~~&$[bq]_{6}[\bar{q}\bar{q}]_{\bar{6}}$   ~~~&  ~~~&6317 ~~~&~~~  \\
                   &~~             &~~      &~~~   &
                   ~~~&&&&444 ~~~~&$[bq]_{6}[\bar{q}\bar{q}]_{\bar{6}}$  ~~~&  ~~~&6410 ~~~&~~~  \\ 
                   &~~             &~~      &~~~
                   ~~~&&&&&~~ &~~     &~~~    \\ \hline

$IJ^{P}$ ~~~&$F_{i}S_{j}C_{k}$ ~~~~&Channel ~~~~&$E_{th}$ ~~~~~&$E_{sc}$ ~~~~&$E_{cc}$ ~~~~&$E_{mix}$~~~~&~~$IJ^{P}$ ~~~&$F_{i}S_{j}C_{k}$ ~~~~&Channel ~~~~&$E_{th}$ ~~~~~&$E_{sc}$ ~~~~&$E_{cc}$ ~~~~&$E_{mix}$~~\\
$\frac{1}{2}2^{+}$ ~~~&261 ~~~~&$\omega B^{*}$ ~~~&6021 ~~~~&6024 ~~~&6024 ~~~&6024 ~~~&~~$\frac{3}{2}0^{+}$ ~~~&311 ~~~~&$\pi B$ ~~~&5419 ~~~~&5422 ~~~&5422 ~~~&5422~ \\
                    &161 ~~~~&$\rho B^{*}$  ~~~&6092    ~~~~&6093
                   ~~~&&&&321 ~~~~&$\rho B^{*}$  ~~&6092  ~~~~&6094 ~~~&~~~   \\

                   &262 ~~~~&$[\omega]_{8}[B]_{8}^{*}$   ~~~&  ~~~~&6345   ~~~&
                   ~~~&&&312 ~~~~&$[\pi]_{8} [B]_{8}$   ~~~&  ~~~~&6371 ~~~&~~~ \\
                   &162 ~~~~&$[\rho]_{8} [B^{*}]_{8}$  ~~~&  ~~~~&6406   ~~~&
                   ~~~&&&322 ~~~~&$[\rho]_{8} [B^{*}]_{8}$   ~~~&  ~~~~&6296 ~~~&~~~ \\

                   &463 ~~~~&$[bq]_{\bar{3}}[\bar{q}\bar{q}]_{3}$ ~~~&  ~~~~&6305   ~~~&6305 ~
                      &&&623 ~~~~&$[bq]_{\bar{3}}[\bar{q}\bar{q}]_{3}$ ~~~&  ~~~~&6252  ~~~&6208 ~~~&~~~ \\
                   &564 ~~~~&$[bq]_{6}[\bar{q}\bar{q}]_{\bar{6}}$  ~~~&  ~~~&6424
                   ~~~&&&&614 ~~~~&$[bq]_{6}[\bar{q}\bar{q}]_{\bar{6}}$  ~~~&   ~~~~&6415 ~~~&~~~\\ 
                      &~~             &~~      &~~~
                      &&&&&~~ &~~     &~~~    \\ \hline

$IJ^{P}$ ~~~&$F_{i}S_{j}C_{k}$ ~~~~&Channel ~~~~&$E_{th}$ ~~~~~&$E_{sc}$ ~~~~&$E_{cc}$ ~~~~&$E_{mix}$~~~~&~~$IJ^{P}$ ~~~&$F_{i}S_{j}C_{k}$ ~~~~&Channel ~~~~&$E_{th}$ ~~~~~&$E_{sc}$ ~~~~&$E_{cc}$ ~~~~&$E_{mix}$~~\\
$\frac{3}{2}1^{+}$ ~~~&331 ~~~~&$\pi B^{*}$ ~~~~&5458 ~~~~&5461 ~~~&5461 ~~~&5461 ~~~&~~$\frac{3}{2}2^{+}$ ~~~&361 ~~~~&$\rho B^{*}$ ~~~&6092 ~~~~&6095   ~~~&6095 ~~~&6095~ \\
                      &341 ~~~~&$\rho B^{~}$  ~~~~&6053  ~~~~&6056
                      ~~~&&&& ~~~&~~     ~        &~~     &~~~    \\
                      &351 ~~~~&$\rho B^{*}$  ~~~~&6092  ~~~~&6094
                      ~~~&&&&&~~             &~~     &~~~ \\

                      &332 ~~~~&$[\pi]_{8}[B^{*}]_{8}$  ~~~&     ~~~~&6363   ~~~&
                      ~~~&&&362 ~~~~&$[\rho]_{8} [B^{*}]_{8}$  ~~~&  ~~~~&6404 ~~~&~~~ \\
                      &342 ~~~~&$[\rho]_{8}[B^{~}]_{8}$  ~~~&  ~~~~&6379  ~~~&
                      ~~~&&&&&~~             &~~     &~~~  \\
                      &352 ~~~~&$[\rho]_{8}[B^{*}]_{8}$  ~~~&   ~~~~&6335 ~~~&
                      ~~~&&&&&~~             &~~     &~~~ \\

                      &633 ~~~~&$[bq]_{\bar{3}}[\bar{q}\bar{q}]_{3}$  ~~~&  ~~~~&6294  ~~~&6222
                      ~~~&&&663 ~~~~&$[bq]_{\bar{3}}[\bar{q}\bar{q}]_{3}$  ~~~&  ~~~~&6369  ~~~&6369 ~~~&~~~ \\
                      &653 ~~~~&$[bq]_{\bar{3}}[\bar{q}\bar{q}]_{3}$  ~~~&  ~~~~&6318
                      ~~~&&&~~ &~~    &~~~ \\

                      &644 ~~~~&$[bq]_{6}[\bar{q}\bar{q}]_{\bar{6}}$  ~~~&  ~~~~&6410
                      ~~~&&&~~ &~~    &~~~ \\

\hline\hline
\end{tabular}
\end{table*}

\begin{table*}[htb]
\caption{\label{ECrms2} Contributions of all potentials to the binding energy (unit: MeV) and root-mean-square distances (unit: fm) in $bq\bar{q}\bar{q}$ system.}
\begin{tabular}{ccccccccccccccccccccc}\hline\hline
 $IJ^{P}$&~~~energy&~~~&Kinetic ~~~&Conf ~~~~&OGE ~~~~&$\pi$ ~~~~~&$\eta$~~~~~&$\sigma$ ~~~~~&B.E ~~~~~&$r_{12}$ ~~~~~&$r_{13}$ ~~~~~~&$r_{14}$ ~~~~~&$r_{23}$ ~~~~~&$r_{24}$ ~~~~~&$r_{34}$~ \\ \hline
$\frac{1}{2}0^{+}$&~~5414&~~~&48.0 ~~~&-2.6 ~~~~&-8.4 ~~~~&-25.2 ~~~~&0.9 ~~~~&-17.7 ~~~~&-5 ~~~~~&1.61 ~~~~&2.18 ~~~~&1.61 ~~~~&1.64 ~~~~&2.25 ~~~~&1.64~ \\
\hline
$\frac{1}{2}1^{+}$&~~5456&~~~&34.5 ~~~&-1.9 ~~~~&-2.9 ~~~~&-19.6 ~~~~&0.6 ~~~~&-14.0 ~~~~&-2 ~~~~~&1.86 ~~~~&2.56 ~~~~&1.86 ~~~~&1.89 ~~~~&2.62 ~~~~&1.89~ \\
\hline\hline
\end{tabular}
\end{table*}

From the numerical results in Table~\ref{energy2}, we can see that for the $b\bar{q}$-$q\bar{q}$ system, there is no bound state below the minimum corresponding threshold for single channel. However, after channel-coupling calculation, we obtain two bound states with binding energies $-5$ MeV in the $IJ^{P}=\frac{1}{2}1^{+}$ system and $-2$ MeV in the $IJ^{P}=\frac{1}{2}1^{+}$ system, respectively.
Moreover, we also calculate the contributions of each terms in the Hamilton and the root-mean-square (rms) distances in the $IJ^{P}=\frac{1}{2}0^{+}$ and $IJ^{P}=\frac{1}{2}1^{+}$ systems, respectively, listed in Table~\ref{ECrms2}. From table~\ref{ECrms2}, we can see that for the $IJ^{P}=\frac{1}{2}0^{+}$ system with the energy of $5414$ MeV, the Confinement, OGE, $\pi$-meson exchange and $\sigma$-meson exchange terms provide attractive interactions, while the kinetic term provide the repulsive interaction. The attraction provided by the $\pi$-meson exchange and $\sigma$-meson exchange counteracts most of the repulsion from the kinetic term. So, the contributions of Goldstone boson exchanges play an important role in the formation of the bound states.
Moreover, the rms distances among the quarks are $1.6$-$2.2$ fm and the main component of this bound state is $\pi B$ ($~97\%$), which indicates that this bound state should be a molecular state. The situation is similar for the $IJ^{P}=\frac{1}{2}1^{+}$ system. The lowest energy is $5456$ MeV with the main component of $\pi B^{*}$ (~$98\%$), and the rms distances among the quarks are $1.8$-$2.6$ fm, showing that this bound state is also a molecular state.
Besides, the real-scaling method is also employed to search for resonance states in the $bq\bar{q}\bar{q}$ system. The results are shown in Figs.~9-14, which show that there is no any genuine resonance state in the $bq\bar{q}\bar{q}$ system.


\begin{figure}[htp]
  \setlength {\abovecaptionskip} {-0.1cm}
  \centering
  \resizebox{0.50\textwidth}{!}{\includegraphics[width=2.0cm,height=1.5cm]{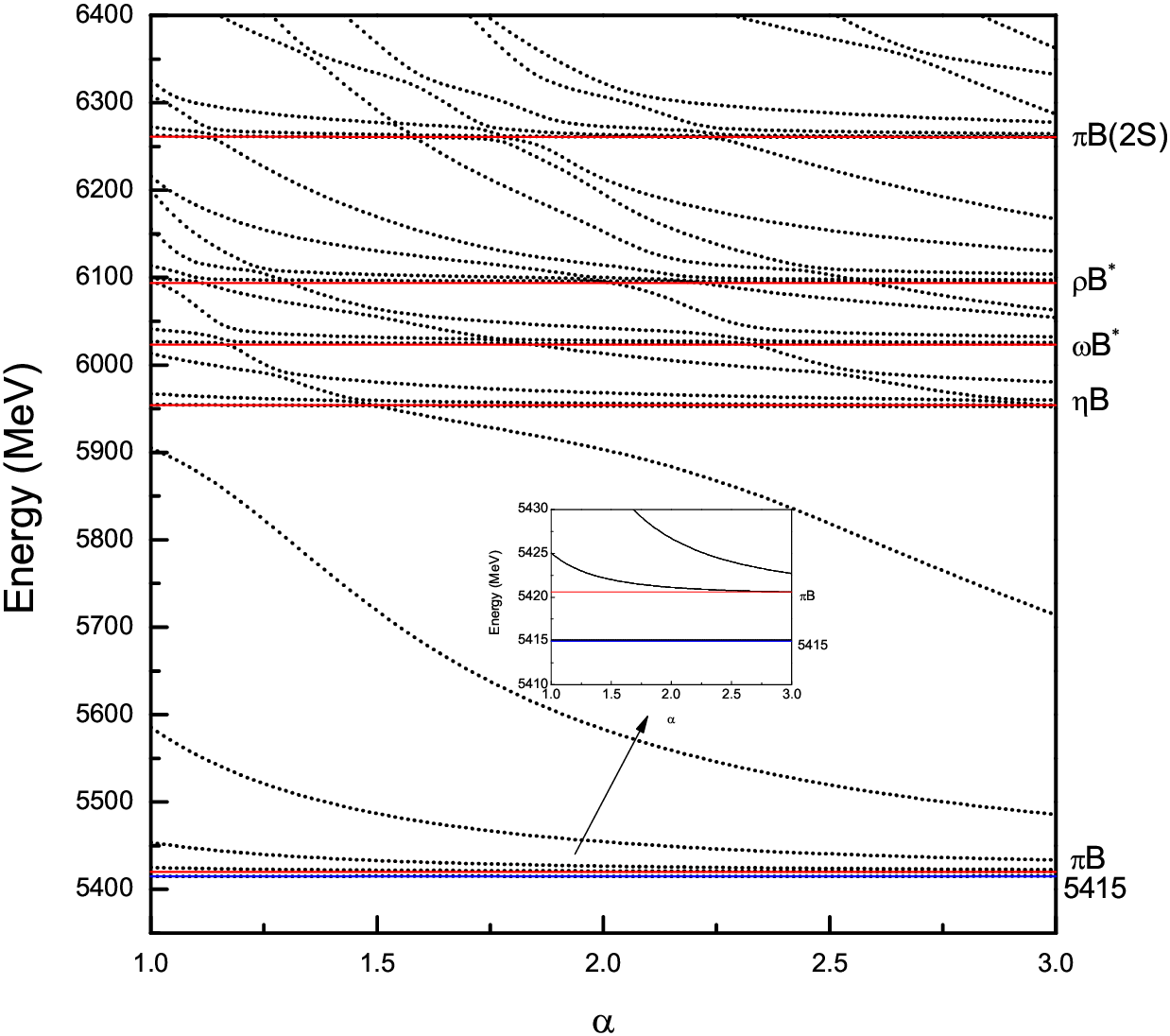}}
  \caption{\label{b1} The stabilization plots of the energies of the $bq\bar{q}\bar{q}$ system with $IJ^{P}=\frac{1}{2}0^{+}$.}
\end{figure}

\begin{figure}[htp]
  \setlength {\abovecaptionskip} {-0.1cm}
  \centering
  \resizebox{0.50\textwidth}{!}{\includegraphics[width=2.0cm,height=1.5cm]{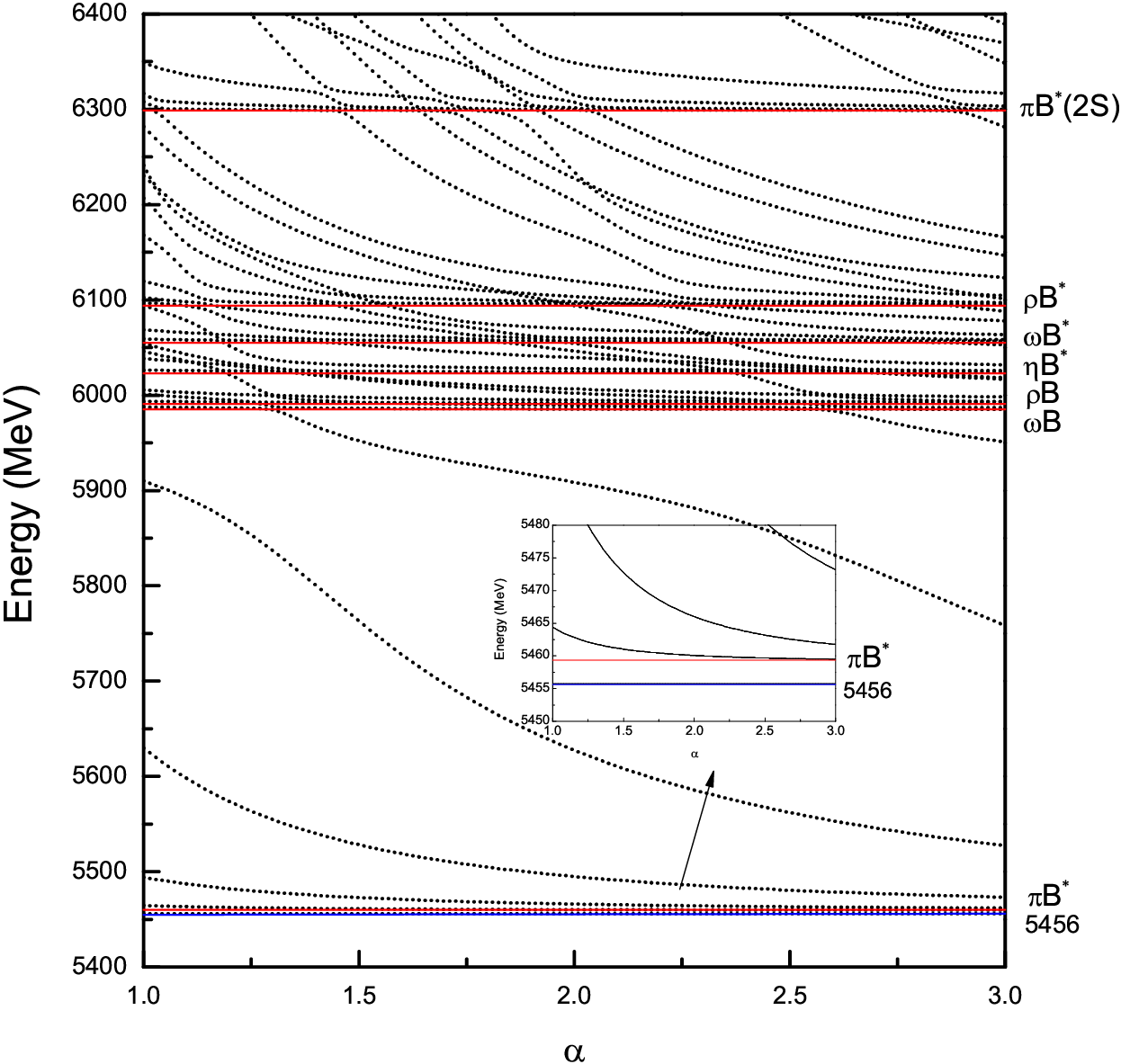}}
  \caption{\label{b2} The stabilization plots of the energies of the $bq\bar{q}\bar{q}$ system with $IJ^{P}=\frac{1}{2}1^{+}$.}
\end{figure}


\begin{figure}[htp]
  \setlength {\abovecaptionskip} {-0.1cm}
  \centering
  \resizebox{0.50\textwidth}{!}{\includegraphics[width=2.0cm,height=1.5cm]{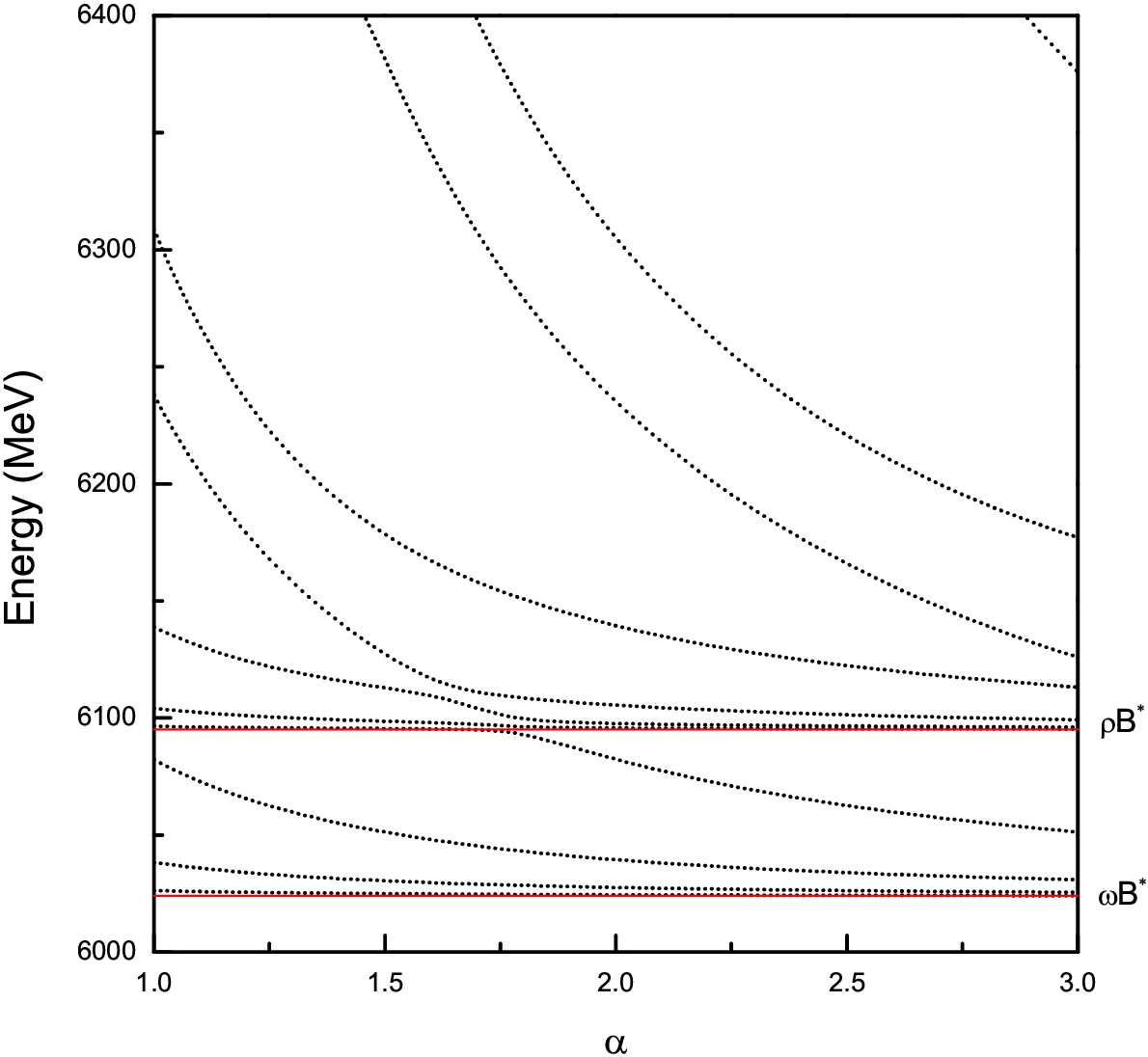}}
  \caption{\label{b3} The stabilization plots of the energies of the $bq\bar{q}\bar{q}$ system with $IJ^{P}=\frac{1}{2}2^{+}$.}
\end{figure}


\begin{figure}[htp]
  \setlength {\abovecaptionskip} {-0.1cm}
  \centering
  \resizebox{0.50\textwidth}{!}{\includegraphics[width=2.0cm,height=1.5cm]{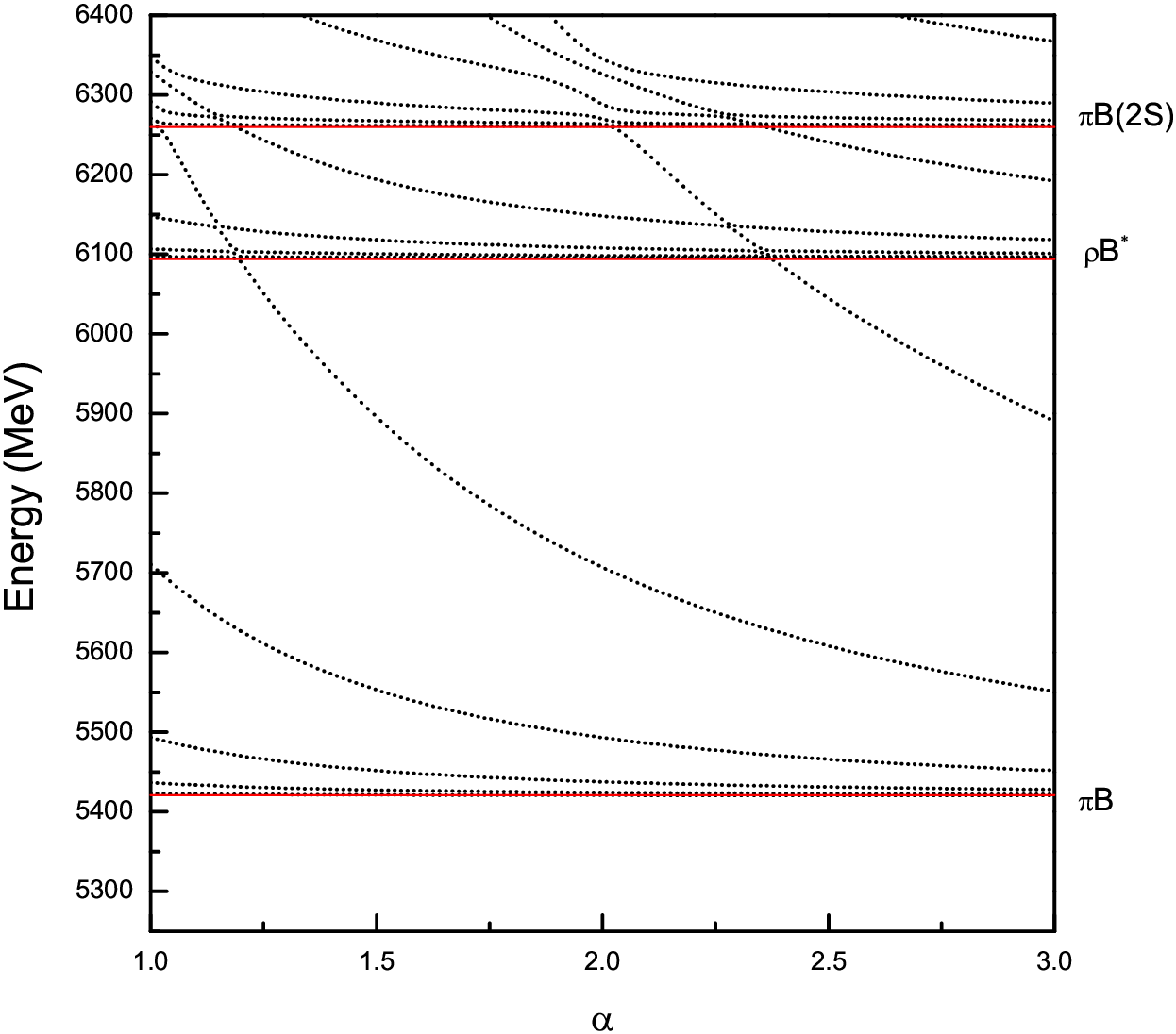}}
  \caption{\label{b4} The stabilization plots of the energies of the $bq\bar{q}\bar{q}$ system with $IJ^{P}=\frac{3}{2}0^{+}$.}
\end{figure}


\begin{figure}[htp]
  \setlength {\abovecaptionskip} {-0.1cm}
  \centering
  \resizebox{0.50\textwidth}{!}{\includegraphics[width=2.0cm,height=1.5cm]{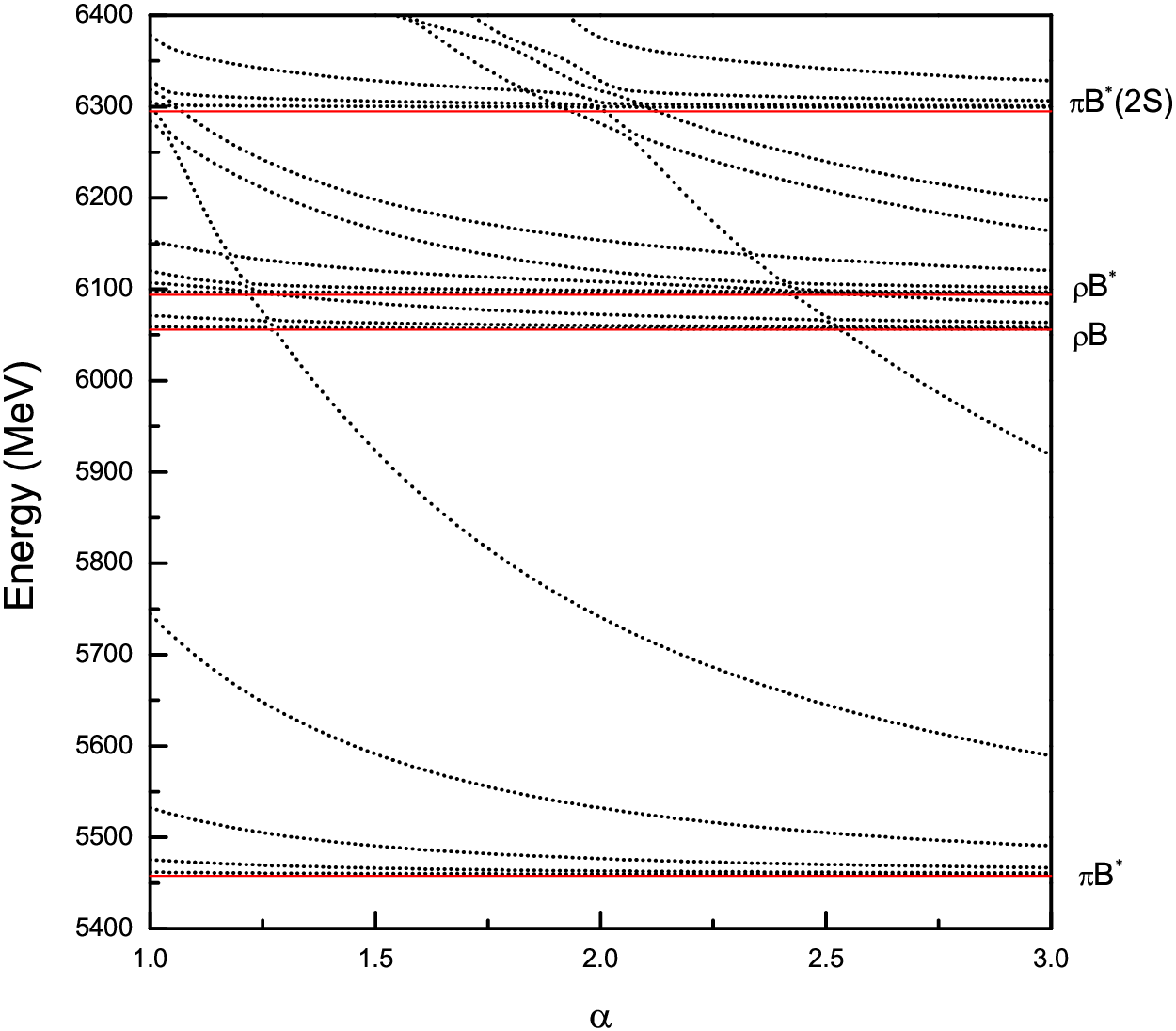}}
  \caption{\label{b5} The stabilization plots of the energies of the $bq\bar{q}\bar{q}$ system with $IJ^{P}=\frac{3}{2}1^{+}$.}
\end{figure}


\begin{figure}[htp]
  \setlength {\abovecaptionskip} {-0.1cm}
  \centering
  \resizebox{0.50\textwidth}{!}{\includegraphics[width=2.0cm,height=1.5cm]{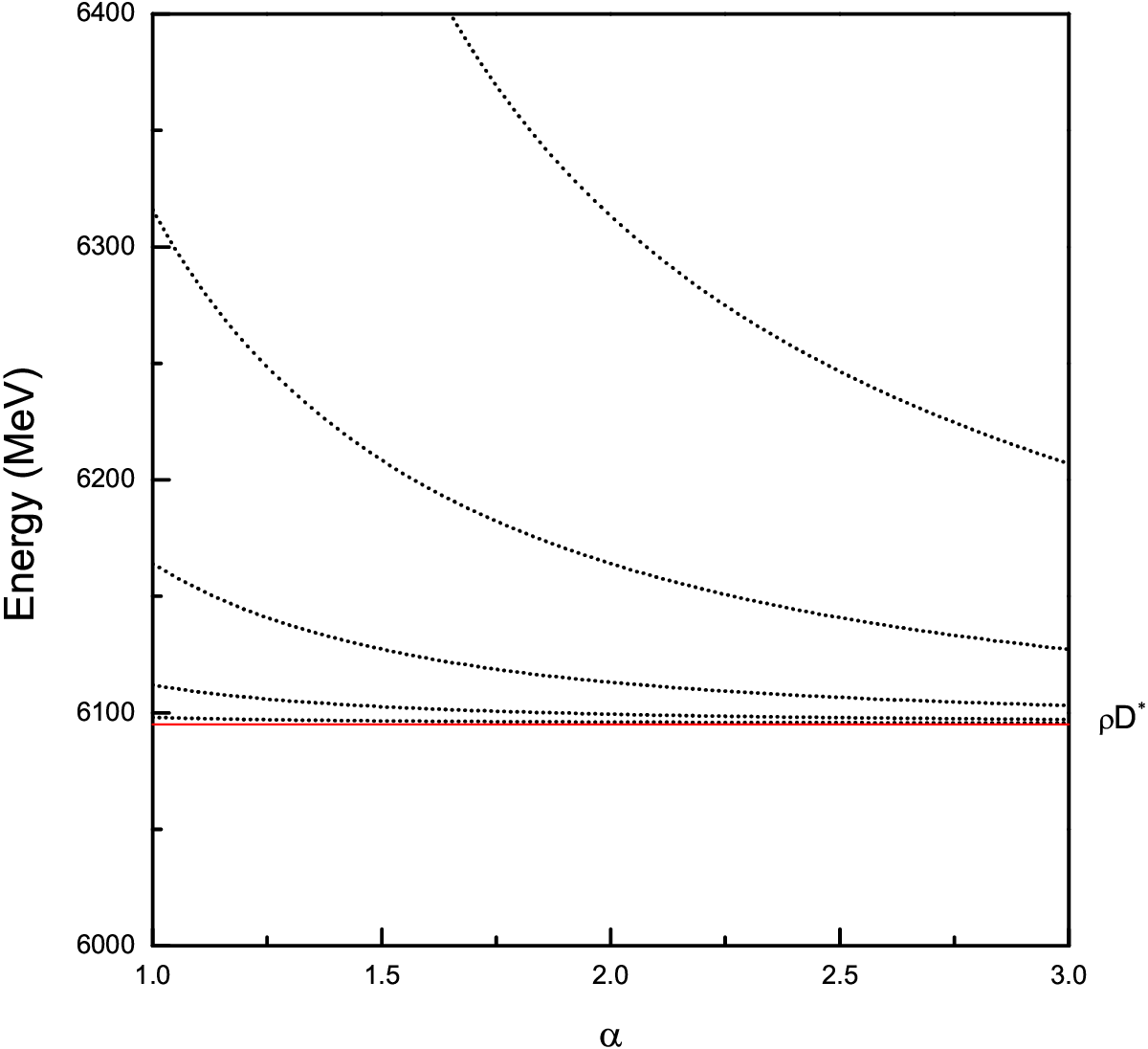}}
  \caption{\label{b6} The stabilization plots of the energies of the $bq\bar{q}\bar{q}$ system with $IJ^{P}=\frac{3}{2}2^{+}$.}
\end{figure}

\section{summary}
In this work, the low-lying system $Qq\bar{q}\bar{q}$ ($Q=c,b$ and $q=u,d$) is systematically investigated in the framework of the ChQM. $Q\bar{q}$-$q\bar{q}$, $Qq$-$\bar{q}\bar{q}$ structures and channel-coupling of these two configurations are considered. In order to search for bound state in the $Qq\bar{q}\bar{q}$ system, dynamical bound-state calculations have been performed. At the same time, a real-scaling method is employed to find the genuine resonance states.

The bound-state calculations show that for the single channel, there is no evidence for any bound state below the minimum threshold in both $cq\bar{q}\bar{q}$ and $bq\bar{q}\bar{q}$ systems.
However, after coupling all channels, we obtain a bound state below the minimum threshold in the $cq\bar{q}\bar{q}$ system with the binding energy of $4$ MeV, and the quantum number is $IJ^{P}=\frac{1}{2}0^{+}$. Meanwhile, in the $bq\bar{q}\bar{q}$ system, two bound states with binding energies of $5$ MeV and $2$ MeV are obtained, and the quantum numbers are $IJ^{P}=\frac{1}{2}0^{+}$ and $IJ^{P}=\frac{1}{2}1^{+}$, respectively.
All three bound states are obtained by channel coupling, suggesting that channel coupling effects are important for the formation of bound states. Moreover,
for these three bound states, we study the contributions of each term in Hamilton, and the results indicate that the Goldstone boson exchanges contributions play a dominant role in the formation of bound states in the $Qq\bar{q}\bar{q}$ ($Q=c, b$ and $q=u, d$) system. To investigate the structure of the bound states, we also calculate the rms distances between quarks and the channel components of the bound states. The results show that all of the bound states are molecular states.

Besides, this work shows that the coupling of various configurations is important to search for resonance states. We consider not only the $cq$-$\bar{q}\bar{q}$ structure, but also $c\bar{q}$-$q\bar{q}$ structure and the channel coupling of the two configurations. It is known that $cq$-$\bar{q}\bar{q}$ states are possible resonance states, but could be coupled to the $c\bar{q}$-$q\bar{q}$ states, such that the $cq$-$\bar{q}\bar{q}$ states may decay to the corresponding threshold. We can estimate whether a resonance state exists after coupling the two structure by employing the real-scaling method. After calculations and analysis with the real-scaling method, we find that the $cq$-$\bar{q}\bar{q}$ states decay to the corresponding threshold. So, there is no genuine resonance state in $cq\bar{q}\bar{q}$ and $bq\bar{q}\bar{q}$ systems in present work.
However, the above three possible bound states are worthy of searching in future experiments.

\acknowledgments{This work is supported partly by the National Science Foundation
of China under Contract Nos. 11675080, 11775118, 11535005 and the Funding for School-Level Research Projects of Yancheng Institute of Technology (No. xjr2022039).

\end{document}